 \documentclass[final,3p,times,twocolumn]{elsarticle}

\usepackage{amsmath}
\usepackage{amssymb}
\usepackage{amsthm}
\usepackage{algorithm}
\usepackage{eurosym}
\usepackage{algpseudocode}

\newtheorem{theorem}{Theorem}
\newtheorem{result}{Result}
\newtheorem{remark}{Remark}
\newtheorem{nono-theorem}{Theorem}[]
\newtheorem{nono-result}{Result}[]

\usepackage{amssymb}
\usepackage{natbib}
\usepackage{comment}

\makeatletter
\def\ps@pprintTitle{%
  \let\@oddhead\@empty
  \let\@evenhead\@empty
  \def\@oddfoot{\reset@font\hfil\thepage\hfil}
  \let\@evenfoot\@oddfoot
}
\makeatother

\begin{document}

\begin{frontmatter}



\title{A fundamental Game Theoretic model and approximate global Nash Equilibria computation for European Spot Power Markets}


\author[inst1]{Ioan Alexandru Puiu}

\affiliation[inst1]{organization={Mathematical Institute, University of Oxford},
            addressline={Andrew Wiles Building, Woodstock Road}, 
            city={Oxford},
            postcode={OX26GG}, 
            state={Oxfordshire},
            country={United Kingdom}}

\author[inst1]{Raphael Andreas Hauser}

\begin{abstract}
Spot electricity markets are considered under a Game-Theoretic framework, where risk averse players submit orders to the market clearing mechanism to maximise their own utility. Consistent with the current practice in Europe, the market clearing mechanism is modelled as a Social Welfare Maximisation problem, with zonal pricing, and we consider inflexible demand, physical constraints of the electricity grid, and capacity-constrained producers. A novel type of non-parametric risk aversion based on a defined worst case scenario is introduced, and this reduces the dimensionality of the strategy variables and ensures boundedness of prices. By leveraging these properties we devise Jacobi and Gauss-Seidel iterative schemes for computation of approximate global Nash Equilibria, which are in contrast to derivative based local equilibria. Our methodology is applied to the real world data of Central Western European (CWE) Spot Market during the 2019-2020 period, and offers a good representation of the historical time series of prices. By also solving for the assumption of truthful bidding, we devise a simple method based on hypothesis testing to infer if and when producers are bidding strategically (instead of truthfully), and we find evidence suggesting that strategic bidding may be fairly pronounced in the CWE region.
\end{abstract}

\begin{keyword}
Spot Electricity Market \sep Game-theoretic Model \sep Numerical Algorithms \sep Risk Aversion \sep Network Constraints \sep Real-world Case Study
\MSC 74S99 \sep 90C20 \sep 91A10 \sep 91B26 
\end{keyword}

\end{frontmatter}


\section{Introduction}\label{sec:intro}
Electricity spot markets throughout the world have undertaken strong deregulation in the last two decades, with the end goal of encouraging investments and increasing economic efficiency \cite{Wilson2002}. Due to the complexity of electricity distribution and difficulty of storage, spot electricity markets are generally structured as Day Ahead Auctions, which are usually defined by a market clearing mechanism that allocates production and consumption quantities, and determines prices. However, even today, many challenges are yet to be overcome, and extensive research was performed to attempt ensuring these Market Clearing Mechanisms (MCMs) satisfy desirable properties. Nevertheless, the negative result of Myerson and Satterwaite \cite{Myersonimpossib} shows, albeit for a greatly simplified case, that the four desirable properties they define cannot be satisfied simultaneously by any mechanism. This reveals the complexity of these markets and indicates that desirable economic properties cannot simply be assumed to hold, but have to be guaranteed through the MCM. In the game theoretic study of \cite{Hobbs2000}, the authors recognise that transmission constraints and market concentration may prevent power markets from being fully competitive. In other words, truthful bidding (that fully reflects cost or benefit structure) is not guaranteed, and participants may place strategic orders. Other properties such as economic efficiency, budget balance, or incentive for producers to participate may also not be guaranteed. As a result of these difficulties, a large number of Market Clearing Mechanisms were devised, according to different policy requirements of different countries. These mechanisms can have widely different rules and properties, and thus theoretical or simulation results from one market cannot be simply generalised to another. Nevertheless, a thorough understanding of each particular market is of great interest for all entities involved in this market: policy makers, market participants grid operators, and researchers. 
\par
To the best of our knowledge, not much attention was given to fundamental game-theoretic studies of market clearing mechanisms that are directly applicable to Day Ahead Auctions for European Power Markets. We here propose a game-theoretic model based on a market clearing mechanism that is consistent with the methodology of European spot power markets, and with European Power Exchange clearing mechanism in particular. By applying our model to the real world case of what was the Central Western European spot electricity market over the period of 2019-2020 (now extended to a 12 country market), we show good representation of the true price series. We further show that based on a simple hypothesis testing approach, the assumption of truthful bidding, which is usually employed by practitioners and researchers for the CWE market, may be greatly overused. This is because for the period considered, we can assert with high confidence that players were bidding strategically for a much larger fraction of times, than we can assert that they were bidding truthfully \cite{hobbs2005}. 
\subsection{Brief literature survey and motivation}
Game theoretic models for electricity markets have been extensively studied. These are often more realistic than purely competitive economic equilibrium models, due to market complexity and the fairly small number of producers in a market. Early competition models applied to spot electricity markets are based on what is known as Cournot \cite{Hobbs_2xCournot} or Bertrand competition. In Cournot competition, it is conjectured that each player assumes that opponents do no change their offered if the player changes the strategy, while Betrand competition conjectures that offered price of opponents is unaffected. However, these models are not realistic as in practice market participants choose neither the offered quantity nor the offered price. Instead, (proposed) price-quantity pairs, referred to as orders, are submitted to the market clearing mechanism, and these are accepted or rejected based on their feasibility and economic merit \cite{Meyer1989}. The strategy of each player is then not the quantity or price offered but rather orders defined by price-quantity pairs. Game theoretic models of this type are often referred to as supply function equilibrium models. 
\par 
We concern ourselves with this type of more realistic models. These are also well investigated in the literature, but are not directly applicable to the European Market Clearing Mechanism, such as EPEX spot. 
This is because most literature studies use a market clearing mechanism based on Optimal Power Flow (OPF) grid modelling, and Locational Marginal Prices (LMP). By contrast, European markets model the grid via the Flow Based Market Coupling (FBMC) approach, which are explained in detail in \cite{JAODocumentation}, \cite{WPEN2014}, and \cite{WPEN2015}. Further, Zonal Marginal Pricing (ZMP) is used, in which a uniform price for each zone (country) is set according to the highest marginal price of accepted asks. This yields a very different mechanism, and as stated in \cite{Hobbs2000}, many studies show that market peculiarities can have a great impact on market power, and therefore on market outcome.
\par 
Some of the most related research includes the work of \cite{Hobbs2000}, where a game-theoretic oligopolistic model based on a Market Clearing Mechanism with OPF grid model with LMP pricing is considered. Bids and offers are parametrised via linear marginal prices with respect to cleared quantity, and the demand is considered as flexible. A leader-follower (Stackelberg) model is considered, where only one or a few participants manipulate prices, while the rest naively assume that they cannot impact prices. As generally the case with these models, players solve a profit maximisation problem that depends on the MCM outcome, which is itself given by an optimisation problem. This results in bi-level optimisation for each player, that is often reformulated as a Mathematical Program with Equilibrium Constraints (MPEC), and the collection of these for all players results in an Equilibrium Program with Equilibrium Constraints (EPEC) \cite{Hobbs2000}, \cite{DRalph2007}. The increased realism obtained by modelling the MCM and players' strategy space as price-quanitity pairs comes at the expense of increased mathematical complexity and computational cost. A very similar game-theoretic model, based on OPF-LMP market clearing is analysed in \cite{DRalph2007}, where conditions for existence of pure Nash Equilibria are derived, under the same linear parametrisation of orders. Further, diagonalisation algorithms based on Gauss-Seidel and Jacobi approaches are given for numerical computation of equilibria. However, the existence conditions are only proved for the OPF-LMP case. 
\par
It is generally recognized by researchers that solving EPECs is very challenging and computationally intense \cite{bookHobbs}. This can be observed by the fact that real world case studies based on these models are rather scarce or over-simplified. A large part of the computational complexity can be attributed to the Market Clearing Mechanism, which is generally a difficult and large scale optimisation problem, with many and difficult constraints. Certainly, if non-convexities of the market such as start-up costs, ramping times, block orders and others are considered, simply obtaining the global optimum of the MCM problem is extremely challenging as this is a Mixed Integer Programming \cite{hobbs2005}. The extra layer of complexity added by considering strategy adjustment via a game-theoretic framework drastically further increases the computational complexity. Alternative approaches are proposed in \cite{Bruninx2022}, where a Neural Network is used to approximate the MCM, and in \cite{KaracaNoRegret}, where a no-regret algorithm is used to obtain coarse-correlated equilibria in a stochastic environment with partial information. However, these alternative approaches are subject to typical pitfalls of data-driven methods.
\par 
The aforementioned research on game-theoretic models is not directly applicable to European spot electricity markets, and to the best of our knowledge, the literature of such game-theoretic models is very scarce. Further, obtaining numerical solutions to such models is generally very computationally intense, and detailed numerical results on real world case studies are generally not available. The work in this paper aims to address these problems.
\subsection{Main contributions}
 By contrast with the literature, we therefore consider a Market Clearing Mechanism based on FBMC with Zonal Marginal Pricing, but similar to \cite{DRalph2007} and \cite{Hobbs2000}, we use a linear parametrisation of marginal prices as a function of quantity. However, instead of forming an EPEC, which is difficult to solve, we perform (diagonalised) grid search in the strategy space for each player, and solve the MCM exactly via a Convex Quadratic Program formulation. We consider inflexible demand and introduce a novel type of non-parametric risk aversion that bounds and reduces the dimensionality of the strategy space, and thus facilitates efficient grid search. By contrast to the typical EPEC solutions, our approach leverages problem-specific structure to reduce computational effort. Further, solutions to EPECs generally only guarantee weak notions of equilibrium such as local Nash Equilibria, or stationary Nash Points \cite{DRalph2007}. Such concepts are only useful if players are not prepared to drastically change their strategies over limited time periods. However, by contrast, our approach offers a global pure Nash Equilibrium in a discretised strategy space (if such a point exists), and we refer to this as approximate global Nash Equilibrium, since the grid can in principle be refined arbitrarily well. The main contributions of our work are then as following:
 \begin{enumerate}
     \item We introduce a MCM consistent with European spot electricity markets, and model strategic player behaviour under a game-theoretic framework. We devise a computationally efficient approach to obtain approximate global Nash Equilibrium points, if such points exist.
     \item We introduce a novel type of non-parametric risk aversion which we call Robust Strategy Selection (RSS). By contrast to mean-variance risk aversion approaches \cite{mihaRH2015}, RSS bounds the strategy space, ensuring boundedness of prices even under inflexible demand. Further, by contrast to other works such as \cite{weberover1999} and \cite{Hobbs2000}, where dimensionality of strategy space is forcefully reduced by holding one parameter constant, our dimensionality reduction occurs naturally due to the risk aversion. A further advantage of this approach is the non-parametric form that requires no assumption over the distribution of possible outcomes. The properties of RSS facilitate efficient (diagonalised) grid search for equilibrium strategies.
     \item We apply our model to the real world case of Central Western European Day Ahead Auction over the period of 2019-2020 and show that the model offers a good representation of the historical time series of prices. 
     \item Finally, we challenge the un-tested assumption that producers are bidding truthfully in the CWE market. In particular, based on the MCM used in the CWE, there is no theoretical evidence that participants are incentivised to bid truthfully. To this end, we devise a simple method based on hypothesis testing to infer if and when players are bidding strategically instead of truthfully. This approach requires the outputs of our Game Theoretic model and a truthful bidding model which simply solves the MCM based on the true cost structure instead of strategic orders. We show evidence that strategic bidding may be fairly pronounced, in contrast to what researchers, practitioners and policy makers usually assume for the CWE market.
 \end{enumerate}
 \subsection{Discussion of Model features}
 The complexity of our approach is a result of the attempt to realistically model strategic bidding in European Day Ahead Auctions. Admittedly, ignoring non-convexities of power plants and assuming linear marginal prices for orders is a simplification, but these are very often used \cite{DRalph2007}, \cite{Hobbs2000}. However, we do not make other unrealistic assumptions that are commonly employed in the literature such as: (i) price taking, (ii) the ability of players to directly control their accepted quantities, (iii) ignoring network constraints, (iv) ignoring capacity constraints, (v) purely flexible demand, or (vi) using over-simplified pricing schemes. In an oligopoly environment such as spot electricity markets, market players are aware that they influence prices, and the assumption of controlling the traded quantity is simply not true. However, the removal of these two commonly used assumptions greatly increases the difficulty of the problem, as the KKT conditions of both the MCM and producer optimisation problems become Nonlinear Complementarity Problems (NCP). These cannot be easily solved or converted to a Quadratic Program, as is the case for the Linear Complementarity Problems (LCP) resulting when using assumptions (i), (ii), such as in \cite{mihaRH2015}, \cite{Hobbs_2xCournot}, where the combined KKT conditions can be transformed to a Quadratic Program. Assumptions (i) and (ii) thus facilitate theoretical analysis and faster computations at the expense of reduced realism. By contrast, our approach focuses on retaining realistic features and remaining computationally tractable, albeit theoretical analysis of the resulting model is much more challenging, and is only briefly addressed here for a further simplified MCM. This brief theoretical analysis reveals some issues concerning the CWE market clearing mechanism, and motivates the introduction of our novel non-parametric risk aversion.
\subsection{Background information}
We briefly review the concept of auction mechanism, the model of market clearing and network constraints used in the CWE market, the idea of strategic bidding, role of Game Theory in our analysis and the concept of Nash Equilibrium.
\par 
In contrast to typical spot financial markets, spot electricity markets are organised mostly in Day-Ahead Auctions due to the complexity of transmission coupled with difficulty of storage. As a result, Day-Ahead electricity markets only allow for placing limit orders once per day, before the market gets cleared for the next day, once and for all. The market is usually cleared separately, for each 15 minutes or 1 hour interval in the next day, although inter-temporal constraints between these slots are sometimes considered. This type of market clearing is required to ensure balance of supply and demand simultaneously with other market specific feasibility constraints.
\par
A market clearing mechanism then usually has an objective, which could be for example Social Welfare Maximisation \cite{JAODocumentation}, Cost Minimisation, or any other objective usually defined by the regulators with the goal of achieving a desirable economic outcome. Feasibility constraints are also enforced by the MCM, ensuring balance of supply and demand, feasibility of electricity transmission, feasibility of production levels and their changes over time \cite{mihaRH2015}, and others. As a result, any market clearing mechanism can be expressed as a mathematical optimisation problem. This mechanism must include a quantity allocation and a pricing mechanism, and to maintain mathematical simplicity, price determination is usually performed in a second optimisation step, once the first step of determining optimal quantity allocation is complete \cite{ETH2017}, but this does not have to be the case. 
\par
As we shall see in the next section, the CWE market is split in multiple zones (usually countries), each containing many production and consumption nodes, and transmission lines. The corresponding MCM uses what are known as Flow Based Market Coupling (FBMC) network constraints, that are essentially a set of linear constraints applied on the net power exports on each zone $\hat{y}_z$. The linear constraints then have the form $M_p \hat{y} \leq \hat{b}_p$. The matrix $M_p$ is the collection of what are known as Power Transmission Distribution Factors, while the $\hat{b}_p$ vector is obtained using what are known as the Remaining Available Margins (RAMs) \cite{JAODocumentation}. We discuss how the constraint set $(M_p,\hat{b}_p)$ is obtained for the real world case in more detail in Section \ref{sec:numRes}, where we apply our model to the real world case of CWE. However, since available historical data of these constraints is very sparse \cite{PuiuHauser2021}, a complete explanation of how the full set of constraints is obtained for each time index from publicly available data is beyond the scope of this work, but this is the topic of our paper in \cite{PuiuHauser2021}. Nevertheless, for the purpose of this work we only require using that $M_p \hat{y} \leq \hat{b}_p$ is the set of linear constraints ensuring feasibility of transmission, often referred to as network constraints.
\par 
In simple terms, the CWE DAA Market Clearing Mechanism determines the market outcome by first solving a (primal) optimisation problem corresponding to maximising social welfare subject to supply and demand balance, network constrained and other constraints. A pricing problem is then solved. This uses the the dual variables of the primal problem to account for the economic merit of transmission, and a rather arbitrary objective that ensures the price is "intuitive" \cite{JAODocumentation}.
\par 
It is not obvious based on the definition of this mechanism, what is the resulting player behaviour and market outcome. A simple assumption allowing for outcome analysis is that participants bid truthfully, according to their true cost or benefit structure. However, especially in an oligopolistic market such as electricity markets, there may be strong incentives to bid strategically. Game Theory is the standard mathematical tool for fundamental analysis of (fully rational) strategic behaviour. In this framework, each player $i$ has an utility $u_i: \mathcal{S}_ i\to \mathbb{R} $ to maximise, over a strategy space $\mathcal{S}_i$, given other players' strategies denoted as $\sigma_{-i} \in \mathcal{S}_{-i}$. Perhaps the single most important concept is that of a Nash Equilibrium, which is a strategy set $\sigma^*=\{\sigma_1^*,..., \sigma_i^*,...\sigma_n^* \}$ such that no player can unilaterally deviate and increase their utility, that is, for every player $i$ we have that
\begin{equation}
     u_i(\sigma_i^*; \sigma_{-i}^*) \geq u_i(\sigma_i; \sigma_{-i}^*), \hspace{1mm} \forall\sigma_i \in \mathcal{S}_i.
\end{equation}
In auction design, market clearing mechanisms are then analysed under a game-theoretic framework to verify if the MCM attains desirable economic properties. By contrast, while we provide some simple analysis of interest to auction design, the focus of our paper is on fundamental estimation of market outcomes when considering strategic behaviour, and the application of the resulting game-theoretic model to the real world case of CWE market. To our knowledge, this was not done before to the same level of detail on real world data.
\par
The rest of the paper is structured as follows. Section 2 introduces our proposed model representing the CWE market clearing while Section 3 is concerned with the strategic behaviour of market participants under a game-theoretic framework. In Section 3 we further reveal some (undesirable) theoretical properties of the CWE clearing mechanism, and propose a novel type of player risk aversion. In Section 4 we present numerical approaches for obtaining approximate global Nash Equilibria, while in Section 5 we are concerned with the application of our game-theoretic model to the real world case of CWE spot power market. 
\section{Social Welfare Maximisation for Market Clearing}
We now present a market clearing model initially introduced by us in \cite{PuiuHauser2022P1}, that models the practice of market clearing procedure in the Central Western European (CWE) power market. To our knowledge the analysis and application of optimisation models directly reflecting the CWE market is scarce in the literature.
\par 
We consider multiple zones $z \in \mathcal{Z}$, with $\mathcal{P}_z$ producers per zone. We take the demand in each zone denoted as $d_z$ as inflexible, which facilitates efficient computation of the zonal prices and avoids the issue of paradoxically accepted orders \cite{PuiuHauser2022P1}. To maintain computational tractability we neglect block order bids, and similarly to \cite{DRalph2007}, \cite{Hobbs2000} and others, we assume that producers bid linear marginal ask prices $\lambda_i$, at sold quantity $x_i$, of the form
\begin{equation}
    \lambda_i(x_i)=m_ix_i+a_i,
\end{equation}
where $m_i>0$ and $a_i>0$ are order constants. This takes the same form as their cost structure, assumed to be $C_i(x_i)$ given by
\begin{equation}
    \frac{\partial}{\partial x_i}C_i(x_i) = c_ix_i + b_i.
\end{equation}
Considering inflexible demand and the form of the bids we assumed, the social welfare objective to be maximised becomes
\begin{equation}
\begin{aligned}
    F_{SWM} &= -\sum_i \int_0^{x_i} (m_i\chi_i+a_i)d\chi_i \\ &= -\sum_{i}\left[\frac{1}{2}m_i x_i^2+a_i x_i\right]\\
    &= -\frac{1}{2}x^TD_mx-a^Tx,
\end{aligned}
\end{equation}
where $D_m$ is a diagonal matrix with $(D_m)_{i,i}=m_i$. We consider supply and demand balance constraints, capacity constraints and network constraints. Let $x$ be the vector of allocated productions s.t. $x_i$ is the production allocated to player $i$. The supply and demand balance then becomes $1^Tx = d$ where $d=\sum_{z \in \mathcal{Z}}d_z$ is the total demand. We do not allow any producer $i$ to be net short or exceed its production capacity $Q_i$, and therefore we have that each $x_i$ is bounded by $0 \leq x_i \leq Q_i$. 

Finally, we consider network constraints. Let $E \in \mathbb{R}^{|\mathcal{Z}| \times |\mathcal{P}|}$ be the matrix that maps production quantities $x$ to zonal production quantities vector $y$, where $y_z= \sum_{i \in \mathcal{P}_z}x_i$ is the total production of zone $z$. Thus $y=Ex$ and further, let $d^Z$ be the zonal production vector, such that $d^Z_z=d_z$. Since the network constraints are imposed for the net production at zonal level, we have that the network constraints are given by
\begin{equation}
    M_p(y-d^z) \leq \hat{b}_p,
\end{equation}
which can be re-written as
\begin{equation}
    M_pEx \leq \hat{b}_p + M_pd^z =:b_p.
\end{equation}
By collecting all the constraints and using that $\min_w f(w)= \max_w -f(w)$, the SWM optimisation problem can then be written as 
\begin{equation}
    \centering
    \begin{aligned}
         \min_{x}&\sum_i\left(\frac{1}{2}m_i x_i^2+a_i x_i\right) = \frac{1}{2}x^TD_mx+a^Tx\\
        s.t.&\hspace{2mm}  M_pEx \leq b_p\\
        &\hspace{2mm}x_i \leq Q_i, \hspace{1mm} \forall i\\
        &\hspace{2mm}x_i \geq 0, \hspace{3mm} \forall i\\
        &\hspace{2mm}1^Tx = d.
    \end{aligned}
\tag{SWM}\label{SWM}
\end{equation}
Note that there are no price variables in this problem, and for inflexible demand, these can be computed once the optimal solution $x^*$ is obtained, as \cite{PuiuHauser2022P1}:
\begin{equation}
    v_z = \max_{k\in \mathcal{P}_z} \{m_kx_k^*+a_k:x_k^*>0\}
\end{equation}
where $v_z^*$ is the price in zone $z$ at optimality of \eqref{SWM}.
Fortunately, since $m_i>0$ $\forall i$ and the constraints are linear, our SWM model is strictly convex quadratic program (CQP), and this can be solved very effectively numerically. We solve the \eqref{SWM} optimisation problem via the \textit{quadprog} package in \textit{Python 3}. 

Under the particular case of one zone, and by ignoring the network constraints the market clearing given by the Social Welfare Maximisaiton problem becomes
\begin{equation}
\begin{aligned}
    \min_x &\sum_{i \in \mathcal{P}}\frac{1}{2}m_ix_i^2 +a_ix_i,\\
    s.t. & \hspace{1mm} \sum_{i \in \mathcal{P}}x_i = d; \hspace{3mm} x_i \geq 0 \hspace{1mm} \forall i .
    \end{aligned}\tag{SWM-s}\label{SWM-s0}
\end{equation}
In the next section we consider strategic bidding and we analyse some of the theoretical properties of the simplified \eqref{SWM-s0}.
\section{Strategic Bidding and the Game-Theoretic Model}
In the CWE market, it is often assumed that market participants place orders purely based on their true costs. However, market orders are each participant's choice and need not be given exclusively by the cost structure. Since each player has its own interest, they may strategize accordingly, and thus the market mechanism needs to incentivise players to bid truthfully, but this is not a guaranteed property \cite{Myersonimpossib}. In fact, achieving this and other desirable properties is a challenging auction design task \cite{Myersonimpossib}, \cite{ETH2016}, \cite{ETH2017}. For this reason, considering strategic bidding offers a more descriptive model of the electricity spot market. We consider the case when players bid strategically to maximise their own profit knowing that the market is cleared by \eqref{SWM}. We assume complete information, that is, each player knows the true values of $c$, $b$, $M_p$, $b_p$ and $d$, and focus on one time period only.
\par
We next introduce producer's optimisation problem given by strategic bidding in subsection \ref{Produceropt}, obtain some theoretical results on a simplified clearing model in subsection \ref{threz}, and introduce a novel non-parametric risk aversion in subsection \ref{sec:RSS}.
\subsection{Producer's optimisation problem}\label{Produceropt}
Each producer submits a supply curve given by $\sigma_i=(m_i,a_i)$, but we assume that the true capacity $Q_i$ of each producer is common knowledge. Each producer aims to maximise its own profit, which can be expressed as the total revenue minus the total cost:
\begin{equation}\label{profit1}
    \pi_i(\sigma_i; \sigma_{-i}) = v_{z_i} x_i - \frac{1}{2}c_ix_i^2 - b_ix_i, \hspace{1mm} \forall i \in \mathcal{P},
\end{equation}
where $z_i:=\{z:i \in \mathcal{P}_z\}$ is the zone of player $i$.
This is subject to production constraints $0 \leq x_i \leq Q_i$, but can be ignored here, since this is enforced at auction level in \eqref{SWM}, and the players do not directly choose $x_i$. The optimisation problem then becomes
\begin{equation}\label{playeropt}
    \begin{aligned}
     \max_{\sigma_i=(m_i,a_i)} &\pi_i(\sigma_i; \sigma_{-i}) = v_{z_i} x_i - \frac{1}{2}c_ix_i^2 - b_ix_i, \hspace{1mm} \forall i \in \mathcal{P}
    \end{aligned}.
\end{equation}
\par
Note that the strategy variables $\sigma_i$ do not explicity appear in the $\pi_i(\sigma_i; \sigma_{-i})$ expression \eqref{profit1}, but they change $\pi_i$ since both $v_{z_i}$ and $x_i$ are functions of $\sigma_i$ (and $\sigma_{-i}$), as defined by \eqref{SWM}. To estimate their profit for a chosen strategy $\sigma_i$, each player then has to estimate the strategies of all other players, $\sigma_{-i}$, their capacity constraints $Q_{-i}$, the network constraints $(M_p,b_p)$ and demand level $d$, and solve the market clearing problem \eqref{SWM}. However, estimating other players' strategies is particularly challenging especially since historical bids data is not available. Game theory then provides a way for such an estimation, by assuming that each player is perfectly rational and maximises their own profit functions, and knowing that all other players behave identically.
\par We can now see that solving the game theoretic model is a bilevel optimisation problem in which each player solves \eqref{SWM} once for each new strategy set considered, to obtain $v_z$ and $x_i$ and thus a profit value, and the outer optimisation problem \eqref{playeropt} requires the maximum value of this profit. A common way to solve this model is to write the KKT conditions of \eqref{SWM} and combine them with the KKT conditions of \eqref{playeropt} for each player $i$. This results in what is known as an Equilibrium Problem with Equilibrium Constraints (EPEC), but is generally hard to solve \cite{Hobbs2000}, \cite{DRalph2007}. 
\par
However, by leveraging the structure of our problem and the risk aversion introduced in subsection \ref{sec:RSS}, we propose a more computationally efficient method in Section \ref{sec:numRes}, that also gives approximate global NE instead of local stationary Nash points obtained by solving EPECs \cite{DRalph2007}.
\subsection{Characterisation of Equilibria for simplified MCM}\label{threz}
We next show that for a simplified case of a one zone market with no capacity constraints, if producers are perfectly rational and aim to maximise their own profit, there exist infinitely many local Nash equilibria, if demand is inflexible and full information is available. Further, we argue that equilibrium prices could be arbitrarily high.
\par 
Intuitively, this may occur because the mapping from the strategy set to the market outcome given by \eqref{SWM} is not injective, and in particular for a player $i$, multiple strategies may yield the same profit level meaning that the player is indifferent to the choice between these strategies. The next results reveal some characteristics of the equilibria.
\begin{result}\label{unconsSWMresult}
Consider a one zone market with inflexible demand $d>0$ and producer set $\mathcal{P}$. Assume producers have no capacity constraints and that there are no network constraints. Assume that the submitted orders have $m_i>0$, $\forall i$, and that the market clearing takes the form \eqref{SWM-s0}. Then the optimal quantities allocated to active players in set $\mathcal{P}^A$ are
\begin{equation}\label{xivaluncons}
\begin{aligned}
    x_i^* &= \frac{d+ \sum_{j \in \mathcal{P}^A \setminus \{i\}}m_j^{-1}(a_j-a_i)}{m_i \sum_{j\in \mathcal{P}^A}\frac{1}{m_j}}, \hspace{2mm} i \in \mathcal{P}^A\\
\end{aligned},
\end{equation}
and $x_i=0$ if $i \notin \mathcal{P}^A$. Further, the marginal prices for any player in the active set $\mathcal{P}^A$ gives the zonal price, that is $v^*= \lambda_i=m_ix_i^*+a_i$.
\begin{proof}
See \ref{sec:appendix_proofs} for the full proof.
\end{proof}
\end{result}
This result gives us a way to track the impact of each player's strategy on their (and others') quantity allocation. We next show that if $m$ is fixed and known by all players, strategising over $a$ only is well behaved, yielding a unique equilibrium.
\begin{theorem}\label{unique_NE_in_A}
Consider a one zone market with inflexible demand $d>0$ and active producer set $\mathcal{P}$. Assume producers have no capacity constraints and that there are no network constraints. Further assume that demand $d$ and cost structure $\{c_i,b_i\}_{i \in \mathcal{P}}$ is common knowledge, and that each producer is aiming to maximise its profit, $\pi_i(m_i,a_i,m_{-i},a_{-i})=vx_i-\frac{1}{2}c_ix_i^2-b_ix_i$, with $v$ the zonal clearing price. Let the market clearing be given by \eqref{SWM-s0}. Finally, assume that for every $i$ we have $c_i>0$, $m_i^+\geq\frac{1}{2}c_i$ is held fixed and that $a_i$ is the only strategy variable, and that this is common knowledge. Then for any vector $m^+ \in \mathcal{M}$ there exists a unique Nash equilibrium $a^+(m^+) \in \mathcal{A}$, which is defined by the (full rank) linear system
\begin{equation}\label{linsysa}
    \theta_i\sum_{j \in \mathcal{P}\setminus\{i\}}\frac{a_i^+}{m_j} - \sum_{j \in \mathcal{P}\setminus\{i\}}\frac{a_j^+}{m_j} = d + (\theta_i-1)\sum_{j \in \mathcal{P}\setminus\{i\}}\frac{b_j}{m_j},
\end{equation}
where $ \theta_i = \frac{2 - k_iK_{-i}^2}{1 - k_iK_{-i}^2}$, $k_i=2m_i-c_i$, $K_{-i}^2 = \frac{\sum_{j \in \mathcal{P}\setminus\{i\}}m_j^{-1}}{m_i\sum_{j \in \mathcal{P}}m_j^{-1}}$.
\begin{proof}
See \ref{sec:appendix_proofs} for the full proof.
\end{proof}
\end{theorem}
\begin{remark}\label{rem1}
The existence of a unique equilibria is beneficial for computation of the market outcome, but does not imply that the market outcome is purely competitive. In other words, it may be the case that even when $m^+=c$, $a_i^+>b_i$ for some or all $i$, and that $v^*>v_0$, where $v_0$ is the price when players bid $(c,b)$. In fact, numerical simulations suggest that this is generally the case.
\end{remark}
Although the result of Theorem \ref{unique_NE_in_A} may appear to be a positive result in isolation, as we show next, in combination with allowing strategising over $m$, an unsatisfactory result characterises \eqref{SWM-s0}:  infinitely many equilibria can be found, and potentially yielding arbitrarily high prices. Our results are consistent with \cite{Hobbs2000} that observes that \textit{"multiple equilibria are likely"} under $(m,a)$ parametrisation, and with \cite{rossbaldick} that states for the same case that \textit{"multiple Nash Equilibria are almost ineviatble"}. We quantify and prove this in Theorem \ref{Theorem_inf_NE}.
\begin{theorem}\label{Theorem_inf_NE}
Consider identical conditions to Theorem \ref{unique_NE_in_A}, with the exception that $m_i$ are not held fixed, but rather part of the strategy choice for each player $i$. Then for each $m^+ \in \mathcal{M}$, there exists an $a^+(m^+)$, computed according to Theorem \ref{unique_NE_in_A} by taking $m^+$ as fixed, such that $(m^+,a^+)$ is a local Nash equilibrium in the $(m,a) \in \mathcal{M} \times \mathcal{A}$ space. In other words, since $m^+$ is chosen from an infinite set, there are infinitely many local equilibria. Further, the local equilibria form a continuous subspace $\mathcal{MA}^+$.
\begin{proof}
See \ref{sec:appendix_proofs} for the full proof.
\end{proof}
\end{theorem}
\begin{remark}
Theorem \ref{Theorem_inf_NE} suggests that by considering $m$ values as part of the strategy, competition between players does not increase. However, the extra parameter of choice may improve the ability to influence the market, potentially resulting in less competitive outcomes.
\end{remark}
\begin{remark}
While these results are proved for a very simplistic case of one zone, the same result could be expected for multiple zones, since when zones become decoupled, they can be treated individually.
\end{remark}
\begin{remark}
By continuity of the local Nash Equilibria, players can gradually change their strategies along these equilibria by small increments, and eventually reach an equilibrium that is most desirable for everyone, if such a point exists. 
\end{remark}
The infinitely many equilibria are clearly a problem for the purpose of outcome estimation and for ensuring high social welfare and reasonable prices. We further speculate that in this equilibria subspace ($\mathcal{MA}^+$), there may exist points $(m^+,a^+)$ that give arbitrarily high zonal clearing price $v$. However, this is hard to prove as values in $a^+$ become negative for increasingly large $m^+$, making it difficult to place bounds on $\lambda_i = m_i^+x_i^*+a_i^+$. Nevertheless, via numerical experimentation we observed cases where $v$ at equilibrium point $\left(m(k), a^+(m(k)) \right)$ increases linearly with increasing $k$ defining $m(k) = k c$. This problem is expected, since Theorem \ref{Theorem_inf_NE} essentially tells us that there is too much freedom in the strategy choice, and the price is not invariant to these choices. This makes the market un-competitive, and players can use this freedom to push the clearing price towards higher levels. This problem is generally ignored in the literature, and often $m$ is taken as fixed, for example in the works of \cite{DRalph2007}, \cite{Hobbs2000}, \cite{weberover1999} or other artificial parametrisations are considered, as discussed in \cite{rossbaldick}.

In \cite{weberover1999}, it is implied that the choice of parametrisation is not critical to the outcome, because each player $i$ can perform trade-offs between slope and intercept and obtain the same optimal outcome. However this is not sufficient to ensure outcome invariance to parametrisation choice between $a$ and $(m,a)$. We can see by comparing the results of Theorem \ref{unique_NE_in_A} and Theorem \ref{Theorem_inf_NE}, that the choice of parametrisation is indeed critical. The reason for the difference between $a$-parametrisation and $(m,a)$-parametrisation is subtle but important: while it is true that given a fixed strategy $\sigma_{-i} = (m_{-i}^+,a_{-i}^+)$ of opponents and an optimal response $\sigma_{i}^+$, changing the strategy to any other optimal response $\sigma_{i}'$ will not change the profit, as shown in the proof of Theorem \ref{nono-Theorem_inf_NE}, simultaneous change of $\sigma$ for \textit{all players}, can result in an equilibrium with different market outcome. The existence of an infinite subset of optimal responses facilitates the existence of multiple equilibria and also provides a way to alter these equilibria over time in practice.
\par
However, all this analysis assumes availability of perfect information, and risk neutral players. In reality, players are likely risk averse and only have incomplete information. To maintain computational tractibility we do not consider incomplete information here. Further, mixed (stochastic) strategies \cite{Maschler} are also not considered for the same reason. Note that even under perfect information, for the $(m,a)$-parametrisation, each player $i$ is still presented with a dilemma: \textit{which equilibrium strategy will the other players play}. Numerical experimentation reveals that if different players simultaneously play different equilibrium strategies, negative profits for one or even all participants are possible. On the other hand, bidding $m_i \geq \frac{1}{2}c_i$ and $a_i \geq b_i$ guarantees non-negative profit. Thus, in the presence of risk aversion, even under perfect information, acknowledging the possibility of multiple outcomes requires discrimination between previously equivalent best responses. In the next subsection we introduce a sensible approach that rational players could likely employ to circumvent the issue of multiple equilibria. 
\subsection{Risk Aversion: Robust Strategy Selection}\label{sec:RSS}
It is clear that the multiplicity of equilibria comes from the problem that players have \textit{"too much choice"}. This is obvious by contrasting Theorem \ref{unique_NE_in_A} with Theorem \ref{Theorem_inf_NE}. Strategising over $(m,a)$ instead of only $a$ increases the strategy space but does not intensify the competition accordingly. A rational player may recognize that due to multiplicity of equilibria, playing a Nash Equilibria strategy does not guarantee that the outcome will correspond to this particular equilibrium point. Since negative profits are possible when players play different Nash Equilibria strategies, a risk averse player may take extra precautions in selecting the strategy played.
\par 
First, he could observe that playing strategies only of the form $\mathcal{S}_R(i)=\{(m_i,a_i):m_i \geq \frac{1}{2}c_i, \hspace{1mm} a_i \geq b_i \}$, would guarantee that the profit is non-negative as per Result \ref{result_non_neg}
\begin{result}\label{result_non_neg}.
Under the market clearing given by \eqref{SWM}, player $i$ is guaranteed non-negative profit by playing any $\sigma_i \in \mathcal{S}_R'(i):=\{(m_i,a_i):m_i \geq \frac{1}{2}c_i, \hspace{1mm} a_i \geq b_i \}$, for any $(m_{-i}, a_{-i})$.
\begin{proof}
See \ref{sec:appendix_proofs}.
\end{proof}
\end{result}
\par 
Next, the player could assign probabilities to every possible combination of equilibrium strategies that a player may play, and maximise the expected (risk adjusted) profit. However, estimating such probabilities is a very difficult task and wrong estimation could have a detrimental impact on the actual outcome. Thus, instead of considering a mean-variance approach that adjusts the utility function to account for risk, a conservative player may wish to first ensure a minimum profit level in a \textit{worst case scenario}. Note that according to Result \ref{result_non_neg}, such a guarantee with non-negative profit is always possible. If multiple equivalent such strategies exist, then the player may use this choice to maximise his actual profit (as opposed to the worst case). Further, let us assume that all players behave in this way, and that this is common knowledge. The worst case optimisation problem is then
\begin{equation}
    \begin{aligned}
    \max_{\sigma_i \in \mathcal{S}_R(i)}\min_{\sigma_{-i} \in \mathcal{S}_R(-i)} \pi_i(\sigma_i; \sigma_{-i})
    \end{aligned}
    \tag{WCP(i)}\label{WCP},
\end{equation}
and let us call the set of all solutions to \eqref{WCP} $\mathcal{S}_{WCP}(i)$. The advantage of this approach is the lack of parameters or required estimates, at least when assuming complete information. Unfortunately \eqref{WCP} is very difficult to solve since an analytical form for $\pi_i$ as a function of strategy vector $\sigma$ cannot generally be obtained under the \eqref{SWM} market clearing model. Due to \eqref{SWM} yielding the value of $\pi_i$, \eqref{WCP} is a tri-level optimisation problem. Solving this is clearly very challenging. However, players may be willing to simplify \eqref{WCP} to enable for progress, to at least estimate strategies that perform well under the \textit{worst case}. One such simplification would be to consider \eqref{SWM-s0} instead of \eqref{SWM}. While this ignores the capacity and network constraints, as we shall see, it allows for an analytic closed form for the resulting optimal set of strategies $\mathcal{S}_{WCP}(i)$.
\par
Let us now look at how this simplifies the problem. To begin with, under \eqref{SWM-s0} we have a unique price that satisfies
\begin{equation}\label{v_equality_SWM0}
    v=m_ix_i+a_i, \hspace{1mm} \forall i: x_i>0
\end{equation}
By taking derivatives of $\pi_i$ with respect to $a_j$ and $m_j$ for any $j \neq i$ we get that
\begin{equation}
    \begin{aligned}
    \frac{\partial \pi_i}{\partial a_j} =& \left[(2m_i-c_i)x_i+(a_i-b_i) \right]\frac{\partial x_i}{\partial a_j}\\
    \frac{\partial \pi_i}{\partial m_j} =& \left[(2m_i-c_i)x_i+(a_i-b_i) \right]\frac{\partial x_i}{\partial m_j}
    \end{aligned}
\end{equation}
where
\begin{equation}
    \begin{aligned}
    \frac{\partial x_i}{\partial a_j} &= \frac{1}{m_j}\geq 0 \\
    \frac{\partial x_i}{\partial a_j} &= \frac{m_ix_i+a_i-a_j}{m_j^2m_i\sum_k\frac{1}{m_j}}\geq 0
    \end{aligned}
\end{equation}
where the inequality on the second line holds since $m_ix_i+a_i=v=m_jx_j+a_j>a_j$ if player $j$ is active i.e. $x_j>0$. If player $j$ is inactive, then $ \frac{\partial x_i}{\partial a_j}=0$. At the boundary between active and inactive, we have that $a_j=v$. Either way, $\frac{\partial x_i}{\partial m_j}\geq 0$ and  $\frac{\partial x_i}{\partial a_j}\geq 0$, and strict inequalities hold if player $j$ is active. This means that $\pi_i$ is increasing function of $a_j$ and $m_j$, for any $j \neq i$. Thus, the worst case for player $i$, regardless of its strategy is when all other players bid their minimum acceptable values, i.e. $a_j=b_j$ and $m_j = \frac{1}{2}c_j$ for $\forall j \neq i$. This essentially removes the $\min$ optimisation problem from \eqref{WCP}. Further, the \eqref{SWM-s0} optimisation can be solved analytically, and thus we are left with the following maximisation problem for player $i$:
\begin{equation}
    \max_{(m_i,a_i) \in \mathcal{S}_R(i)}\pi_i(m_i,a_i;\frac{1}{2}c_{-i},b_{-i})
\end{equation}
Using Result \ref{unconsSWMresult} we obtain that
\begin{equation}
    \frac{\partial \pi_i}{\partial a_i} = \left[(2m_i-c_i)x_i+(a_i-b_i) \right]\frac{-\sum_{k \neq i}m_k^{-1}}{m_i\sum_km_k^{-1}}+x_i,
\end{equation}
and in the proof of Theorem \ref{Theorem_inf_NE}, we have shown that $\frac{\partial \pi_i}{\partial m_i}=x_i\frac{\partial \pi_i}{\partial a_i}$, and this can be verified by direct calculation. Thus, the optimal strategies form a set, as when setting $\frac{\partial \pi_i}{\partial a_i}=0$ we obtain one equation with two variables, which can be written as
\begin{equation}\label{non-linear_subspace}
    \left[(2m_i-c_i)x_i+a_i-b_i \right]\frac{-\sum_{j \neq i}m_j^{-1}}{m_i\sum_jm_j^{-1}} + x_i= 0, 
\end{equation}
where $x_i$ can be written exclusively in terms of $a$, $m$ and $d$ as per Result \ref{unconsSWMresult}, and $m_j=c_j$ for $j\neq i$. Equation \eqref{non-linear_subspace} gives a non-linear relationship between $a_i$ and $m_i$ that ensures best response under the \textit{worst case} for the simplified \eqref{SWM-s0} market clearing. However, this can be simplified further by noting that the number of players in the real market is generally fairly large, and therefore 
\begin{equation}\label{approx_taking}
    \frac{\partial x_i}{\partial a_i}=-\frac{\sum_{j \neq i}m_j^{-1}}{\sum_jm_j^{-1}}\frac{1}{m_i} \approx -1 \cdot\frac{1}{m_i},
\end{equation}
and this transforms \eqref{non-linear_subspace} to 
\begin{equation}\label{linear_subspace}
    a_i = v^{WCP(i)} - \frac{v^{WCP(i)}-b_i}{c_i}m_i,
\end{equation}
where $v^{WCP(i)} = m_ix_i(m_i,a_i;c_{-i},b_{-i})+a_i$. By using that $x_i = \frac{v-a_i}{m_i}$ to express $x_i$ in $\pi_i =vx_i -\frac{1}{2}c_ix_i^2-b_ix_i$ and assuming price-taking for player $i$, i.e. $\frac{\partial v}{\partial a_i} = 0$ gives the same result as \eqref{linear_subspace}. In fact, approximation \eqref{approx_taking} is perfectly equivalent to price taking. This means that $v^{WCP(i)}$ does not change along the \eqref{linear_subspace} curve, and therefore there is a linear relationship between $a_i$ and $m_i$. The value of $v^{WCP(i)}$ can be computed by fixing any feasible value $m_i^W \geq \frac{1}{2}c_i$ and first computing the resulting $a_i^W$ by equation \eqref{non-linear_subspace} with $x_i$ given as per Result \ref{unconsSWMresult}, with the approximation in \eqref{approx_taking}. The price $v^{WCP(i)}$ can then be computed by solving \eqref{SWM-s0} or using Result \ref{unconsSWMresult}, and since $v^{WCP(i)}$ is constant along \eqref{linear_subspace}, the optimal response strategy set is now fully defined. However, we observed in practice that obtaining $v^{WCP(i)}$ by solving \eqref{SWM} gives better results. Since each player requires solving \eqref{SWM} once, computational cost can be reduced by solving \eqref{SWM} only once for $m=\frac{1}{2}c$, $a=b$, yielding $v^{WCP(i)}=v^{WCP}_z$ to be the same for each player in a zone $z$. We use this approach in practice and therefore in this case the optimal strategy set is given by
\begin{equation}\label{best_response_linear_subset}
 S_{WCP}(i):= \left\{\begin{aligned}
   (m_i,a_i):&a_i = v^{WCP}_{z_i} - \frac{v^{WCP}_{z_i}-b_i}{c_i}m_i,\\ &m_i \geq \frac{1}{2}c_i, \hspace{1mm}a_i \geq b_i 
\end{aligned}\right\}.
\end{equation}
We can clearly see that so long as $v^{WCP}_{z_i} > b_i$, $ S_{WCP}(i)$ is a one dimensional set with bounds $m_i \in [\frac{1}{2}c_i; c_i]$, and $a_i \in [b_i;\frac{1}{2}v_{z_i}^{WCP}+\frac{1}{2}b_i]$ and the choice over set $S_{WCP}(i)$ can be used to maximise actual profit. Note that $a_i=b_i$ and $m_i=c_i$ is a point in $ S_{WCP}(i)$, thus this approach is guaranteed to be superior for player's $i$ profit when compared to simply taking $a_i=b_i$, $m_i=c_i$ because the extra choice can be used to play a game of profit maximisation. 
\par If $v^{WCP}_{z_i} = b_i$, the choice is unique, while if $v^{WCP}_{z_i} < b_i$ the set $ S_{WCP}(i)$ is empty and player $i$ is inactive, but player $i$ can play $(c_i,b_i)$ which ensures that in the case of unexpected outcome the profit is non-negative. Generally, $S_{WCP}(i)$ is a one dimensional set, and since this set is always bounded, we devise a method to compute efficient (approximate) global Nash Equilibria in Section \ref{sec:effNE}. Before presenting the computational approach, we comment on our proposed RSS idea.
\subsection{A comment on Robust Strategy Selection}
The core idea of RSS is to use the extra dimensionality in the strategy space to essentially make player $i$ play two simultaneous games: (i) one of profit maximisation in worst case scenario, and (ii) another of profit maximisation given that all players play game (i). There are clearly many possible choices of such secondary games (or optimisation problems). However, not all these are compatible with the profit maximisation problem (since the resulting system may have no feasible solution). Further, compatible optimisation problems may yield very difficult problems to solve.
\par
Although we make fairly strong assumptions to proceed with RSS, the merit of our approach is finding a compatible objective, representing possible risk aversion behaviour, that can be solved for analytically, and thus reducing the dimensionality of the strategy space for the main game to one (bounded) dimension for each player. This facilitates obtaining approximate global Nash Equilibria instead of purely local ones (or simply stationary points) obtained using EPECs \cite{Hobbs2000}, \cite{DRalph2007}. Our obtained equilibria may not only be much faster to compute, but also more representative of the market due to the global nature of the equilibria.
\par
While admittedly RSS has its limitations, another merit of our work is perhaps more important: we provided a more general idea for research, by trying to use the \textit{excess choice} in strategy to represent risk aversion of players and obtain simplified strategy sets to consider for the main game, instead of arbitrarily reducing the strategy space. 
\section{Numerical Solution and  Equilibria Computation}\label{sec:effNE}
The approach we use for finding Nash Equilibria differs from the one used in \cite{DRalph2007}, where EPECs are solved because: (i) the approach used in \cite{DRalph2007} only finds stationary points (also known as \textit{stationary Nash points}), which are weaker than local Nash Equilibria \cite{DRalph2007}, and here we aim to compute (approximate) global Nash Equilibria, and (ii)  Robust Strategy Selection reduces the strategy space to one (bounded) dimension for each player, making it much cheaper to compute best responses using grid search.
\begin{algorithm}[h!]
\caption{Compute approximate best response function  $\hat{B}_i(m_i;\omega_i,N_{pts}, \mathcal{M})$ under RSS (a-BRF-RSS)}\label{alg:a-BRF-RSS}
\begin{algorithmic}
\Require input $(m_k,a_k, Q_k)$ $\forall k \in \cup_z\mathcal{J}_z \setminus \{i\}$, $(c_i, b_i)$, $N_{pts}$
\Require solution method $\mathcal{M}$ to solve \eqref{SWM}
\If{$b_i> s_{z_i}^{min}$}
\State return $(c_i,b_i)$
\EndIf
\For{$l=1,2,...,N_{pts}$}
\State get player's $i$ strategy $m_i(l) = \frac{1}{2}c_0 + \frac{1}{2}\frac{l-1}{N_{pts}-1}c_0$ 
\State compute $a_i(l)$ using $(m_i(l),a_i(l)) \in \mathcal{S}_{WCP}(i)$
\State solve \eqref{SWM} using $\mathcal{M}$, and obtain $x^*$, $v^*$ and $y^*$
\State compute $\Pi_i(x_i^*,m_i(l),a_i(l),c_i,b_i)$
\EndFor
\State $m_i^B \gets \textrm{arg} \max_{m_i(l); l \in \overline{1,N_{pts}}}\Pi_i(x_i^*,m_i(l),a_i(l),c_i,b_i)$
\State return $(m_i^B, a_i^B)$
\end{algorithmic}
\end{algorithm}
\par 
Instead of computing derivatives and updating an estimate of a local Nash point until convergence, we consider iteratively improving each player's best response function given other players' strategies. Let us denote the best response function for player $i$, given other players' strategies $(m_{-i},a_{-i})$ as $B_i(m_{-i},a_{-i};\omega_i)$ where the problem parameters for player $i$ are $\omega_i =(c_i,b_i,Q,d, M_p, b_p)$. Since $(m_i, a_i) \in \mathcal{S}_{WCP}(i)$, we drop $a_i$ and only denote the strategy of player $i$ by $m_i$, with the implicit corresponding $a_i$ value given by \eqref{linear_subspace}. Further, we have that $\frac{1}{2}c_i \leq m_i \leq c_i$ and therefore the feasible strategy space is bounded when using RSS. Thus, to compute an approximate best response function, $\hat{B}_i(m_i;\omega_i,N_{pts})$ we discretise the domain in $N_{pts}$ equispaced points over $[\frac{1}{2}c_i, c_i]$, and for each strategy point solve \eqref{SWM} using a CQP solver (\textit{quadprog} in Python 3). Algorithm \ref{alg:a-BRF-RSS} formalises the computation of the best response. Note that our idea for computation of (approximate) global Nash Equilibria remains the same regardless of the market clearing mechanism used. 
\par 
Given an algorithm to compute an (approximate) best response function $\hat{B}_i(m_i;\omega_i,N_{pts})$, one can now  obtain an (approximate) global Nash Equilibrium point by iteratively updating strategies until no further improvement is possible for any player. Similar to \cite{DRalph2007}, we propose two types of iterative updates: (i) a Gauss-Seidel type of approach (GS), where each player uses the most recent information about other players' strategies, and (ii) a Jacobi type of approach (JS), where each player is updated simultaneously using the last strategy point with synchronised information. Both and are relevant here since each approach has its strengths and limitations. The main advantage of GS is that the most recent information is always used, potentially requiring fewer iterations, but paralellisation can only be done across $N_{pts}$ different tasks, as each player has to wait for the previous one's best response function to be updated. On the other hand, JS does not make best use of the last available iterations, and thus may require more iterations, but it has the big advantage that it can be parallelised across $(i,l) \in \mathcal{P} \times \overline{1,N_{pts}}$, i.e across both the player set and discretisation points indepedently. 
\begin{algorithm}[h!]
\caption{Compute strategy update by JS (C-JS)}\label{alg:s6-js}
\begin{algorithmic}
\Require input strategies $m_i$ $\forall i$, $N_{pts}$
\Require solution method $\mathcal{M}$ to solve \eqref{SWM}
\For{$(i,l) \in \mathcal{P} \times \{1,2,...,N_{pts} \}$}
\State get player's $i$ strategy $m_i(l) = \frac{1}{2}c_0 + \frac{l-1}{N_{pts}-1}\frac{1}{2}c_0$ 
\State compute $a_i(l)$ using $(m_i(l),a_i(l)) \in \mathcal{S}_{WCP}(i)$
\State solve \eqref{SWM} using $\mathcal{M}$, for $(m_i(l),m_{-i})$ and obtain $x^*$, $v^*$ and $y^*$
\State compute $\Pi_i(x_i^*,m_i(l),a_i(l),c_i,b_i)$
\EndFor
\For{$i \in \{1,2,...N_p\}$}
\State $m_i^B \gets \textrm{arg} \max_{m_i(l); l \in \overline{1,N_{pts}}}\Pi_i(x_i^*,m_i(l),a_i(l),c_i,b_i)$
\EndFor
\State return $\{(m_i^B,a_i^B) \}_{i \in \mathcal{P}}$
\end{algorithmic}
\end{algorithm}
Thus, $N_pN_{pts}$ tasks can be solved simultaneously, instead of just $N_{pts}$ and assuming identical computational time for each task, and sufficiently many cores to perform independent tasks, it can take $N_p$ times more iterations when compared to GS to achieve the same computational time, albeit at a larger energy usage. Performing a detailed numerical analysis is beyond the scope of this paper and we just use this simple analysis to conclude that JS is more appropriate when sufficiently many cores are available, while GS might be preferred when running on a machine with fewer cores ($\leq32$). The ideas of the two approaches for computing a strategy update for all players are formalised in Algorithm \ref{alg:s6-js} and \ref{alg:s6-gs}.
\begin{algorithm}[h!]
\caption{Compute strategy update by GS (C-GS)}\label{alg:s6-gs}
\begin{algorithmic}
\Require initial strategies $m_i$ $\forall i$, $N_{pts}$
\Require market clearing solution method $\mathcal{M}$
\For{$i \in \{1,2,...N_p\}$}
\State compute $(m_i^B,a_i^B)$ using $\hat{B}_i(m_i;\omega_i,N_{pts}, \mathcal{M})$
\State $m_i \gets m_i^B$
\EndFor
\State return $\{(m_i^B,a_i^B) \}_{i \in \mathcal{P}}$
\end{algorithmic}
\end{algorithm}
\par 
Finally, one can now obtain a Nash equilibrium by repeatedly using C-GS or C-JS using the previous strategy output as input, until convergence, as shown in Algorithm \ref{alg:s6-NE}. To perform intensive computations on the real world data in Section \ref{sec:numRes} we use \textit{DescartesLabs}'s cloud computing services, while the less intensive computations are performed on a machine with 8 cores.
\begin{algorithm}[h!]
\caption{Compute Nash Equilibrium Point (CNEP)}\label{alg:s6-NE}
\begin{algorithmic}
\Require initial strategies $m_i^{(0)}$ $\forall i$, $N_{pts}$, maximum number of cycle $N_{max}^C$, and tolerance $\delta_{NE} \geq 0$
\Require market clearing solution method $\mathcal{M}$
\For{$k \in \{1,2,...N_{max}^C\}$}
\State compute $\{(m_i^{(k)},a_i^{(k)}) \}_{i \in \mathcal{P}}$ using C-GS or C-JS using $\{(m_i^{(k-1)},a_i^{(k-1)}) \}_{i \in \mathcal{P}}$ as input strategies
\If{$m_i^{(k-1)} =m_i^{(k)}$ (or $\|m_i^{(k-1)} -m_i^{(k-1)}\|_2 < \delta_{NE}$)}
 \State break
\EndIf
\EndFor
\State return $\{(m_i^{(k)},a_i^{(k)}) \}_{i \in \mathcal{P}}$
\end{algorithmic}
\end{algorithm}
When running CNEP using the cloud services, we presolve by using CNEP with much coarser $N_{pts}$ and rather small $N_{max}^C$. Under the limited runs performed, we found
this approach to give smaller number of iterations performed as well as shorter computational time.
\section{Numerical Results}
In this section we briefly review some numerical experiments based on synthetic data in subsection \ref{numres:synth} to confirm our analytical results and claims in subsection \ref{threz}. The main focus in on the application of the game-theoretic model (and the model assuming truthful bidding) to the real world case of Central Western European power market over the year commencing on 1 January 2019. Further, we challenge the often employed assumption that producers bid truthfully in the CWE. We devise a simple method to uncover strategic bidding and show that this is fairly pronounced.
\newpage
\subsection{Numerical Experiments on Synthetic Data}\label{numres:synth} 
In this subsection we briefly show numerical results for a synthetic problem under the \eqref{SWM-s0} market clearing, that confirm our findings and claims in subsection \ref{threz}. 
\par
\begin{figure}[ht!]
    \centering
    \includegraphics[width=0.99\columnwidth]{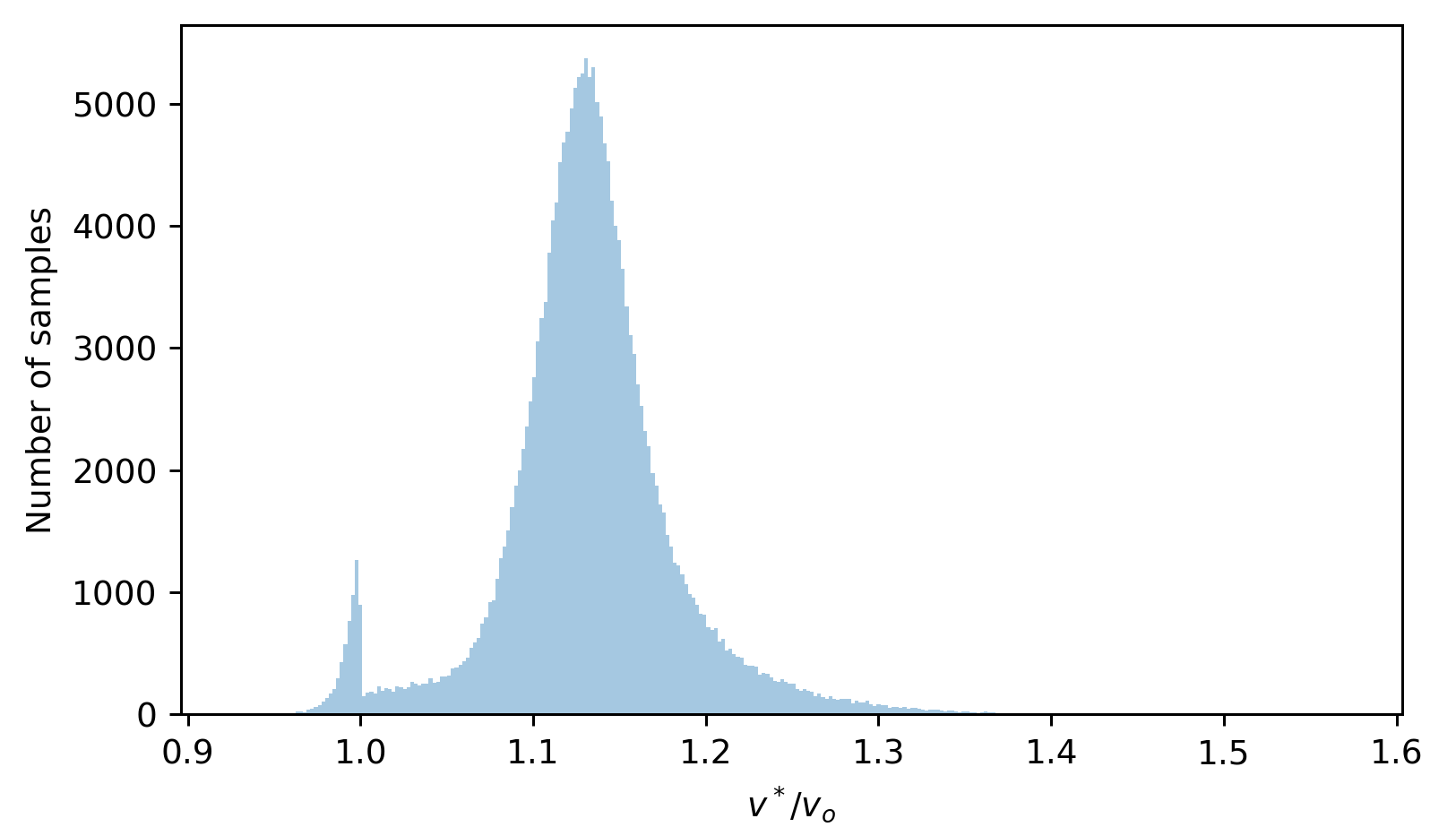}
    \caption{Empirical distribution of $r_v$ ratio for $N=100,000$ samples of $c$ and $b$.}
    \label{fig:dist_ratio}
\end{figure}

 
First, we show that even under the conditions of Theorem \ref{unique_NE_in_A}, the existence and uniqueness of equilibrium does not imply that the outcome is purely competitive, as mentioned in Remark \ref{rem1}. To this end we consider a simulation exercise with one zone and $5$ producers, $d=1$, and sample $c_i \sim U(1,10)$ and $b_i \sim U(0.5,2)$ for $\forall i$ independently. We solve $N=100,000$ such problems and for each problem we solve \eqref{SWM-s0} for two cases: (i) $m_0=c$ and $a_0=b$, and (ii) $m^+=c$, $a^+(m^+)$ given as per Theorem \ref{unique_NE_in_A}, and obtain the resulting prices $v_0$ and $v^*$. We plot the empirical distribution of the ratio $r_v=v^*/v_0$ in Figure \ref{fig:dist_ratio}.
The idea is to reveal how a NE under conditions of Theorem \ref{unique_NE_in_A}, compares with the outcome under truthful bidding. We can clearly see that the price $v^*$ is significantly larger than $v_0$ for the vast majority of times, and the mean of the price ratio is well above $1$. An aware reader may observe that there is a second mode with the mean below and $r_v$ value of $1$. We observe this mode to become more pronounced when $b_i$ values are drawn from a much wider range, when comparing to the range for $c_i$. However, it is clear that under this setup, the outcome of strategising is generally very different than the outcome corresponding to truthful bidding. Further, reasonable value pairs for $(c,b)$ are observed to generally give higher prices than in the case of truthful bidding.
\par
\begin{figure}[ht!]
    \centering
    \includegraphics[width=0.99\columnwidth]{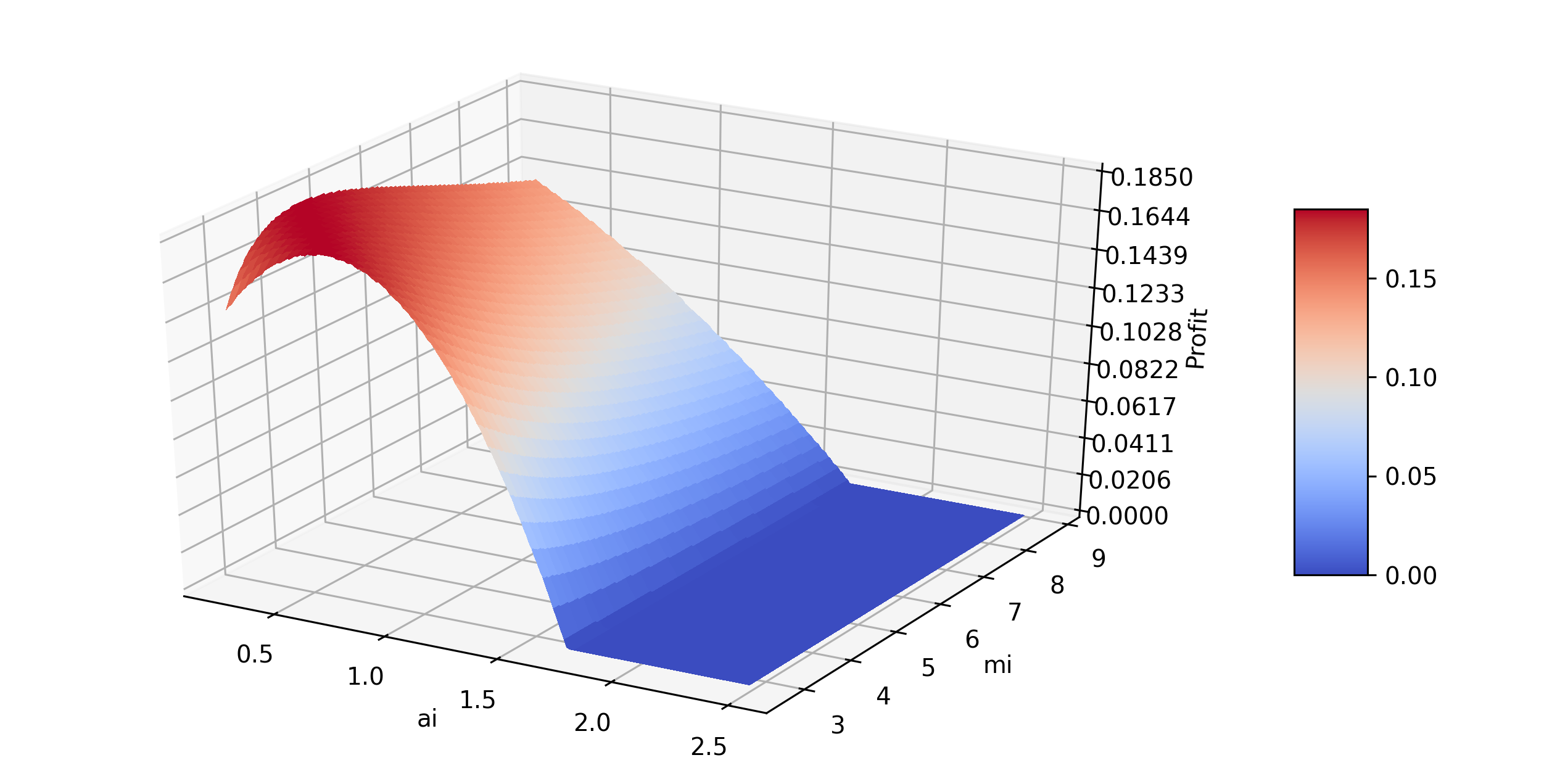}
    \caption{Best response of player $i=0$ for the \eqref{SWM-s0} problem on synthetic data.}
    \label{fig:3D_synth_pi}
\end{figure}

We next consider a specific problem with $c^T=[2.65,1.5,2,1.8,2.2,2.1]$ and $b^T=[0.5,2,1,1.1,0.6, 1]$ as an example, and look at the profit landscape of the first player when all other players bid $m_{-i}=c_{-i}$, $a_{-i}= a_{-i}^+(c_{-i})$ and plot the results in Figure \ref{fig:3D_synth_pi}.
First, we can observe that the profit becomes $0$ for values of $a_i$ above a certain threshold. This is because $x_i^*=0$ is given by \eqref{SWM-s0}. Next, we can clearly see that for a fixed $m_i$, $\pi_i(a_i)$ is a strictly concave function so long as $x_i>0$. We can also see that there exist a range of values $(m_i,a_i)$ that yield the same maximum profit and that $a_i^+ \sim -C_m m_i^+ $ for some $C_m>0$. Finally the plot also confirms our curvature results for derivatives of $\pi_i$ with respect to $m_i$ and provides some intuition as to why multiplicity of equilibria exists.
\par
\begin{figure}[ht!]
    \centering
    \includegraphics[width=0.99\columnwidth]{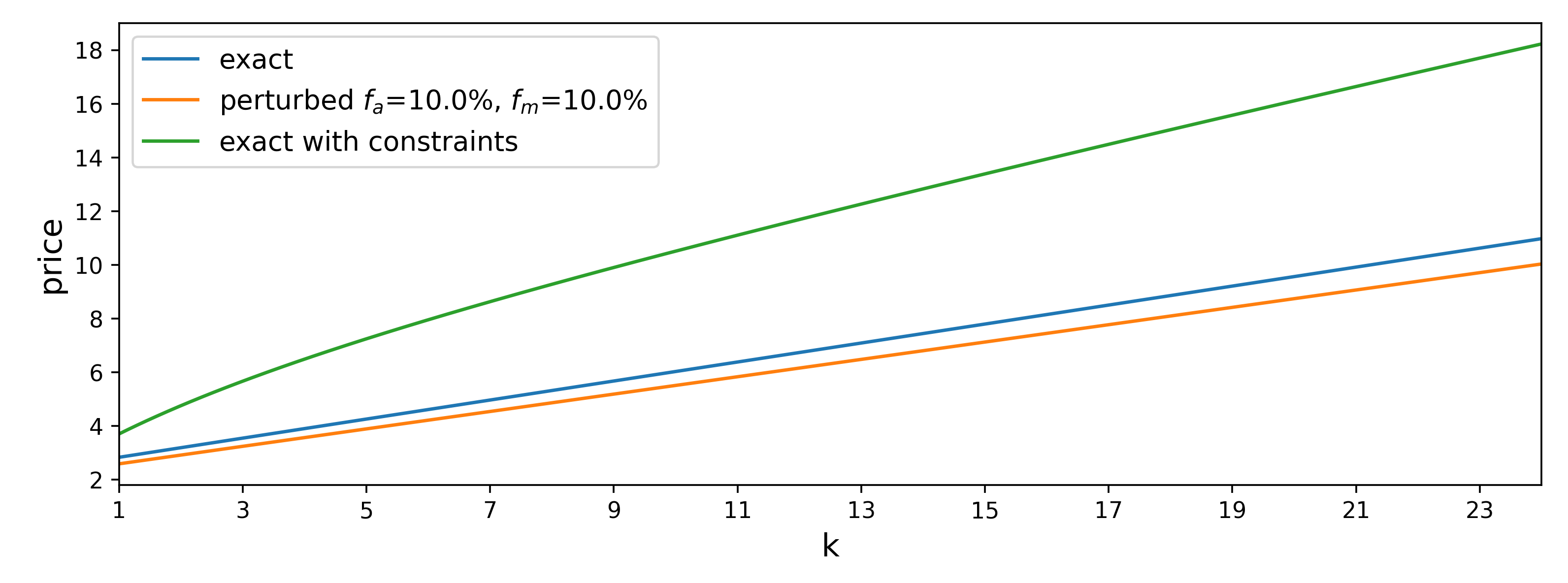}
    \caption{Equilibrium price for the considered synthetic problem with $m(k) = k c$ under \eqref{SWM-s0} (blue and orange curves), and under \eqref{SWM} (green curve).}
    \label{fig:v_synth}
\end{figure}

 
We finally show evidence that equilibrium prices $v^*$ may be arbitrarily large. To this end, we consider $m^+(k) = k \cdot c$ for $k \in [1, K_{max}]$ and for each $m^+(k)$ obtain $a^+(m^+(k))$ as per Theorem \ref{unique_NE_in_A}. We further consider perturbing the strategies for all but player $i=0$ as $\tilde{m}_{-i} = (1-f_m)m^+_{-i}(k)$ $\tilde{a}_{-i} = (1-f_m)a^+_{-i}(m^+(k))$. Finally, we also consider the case of allocating each two consecutive players to a different zone, imposing capacity and network constraints as per the synthetic example in \cite{PuiuHauser2022P1} and solving \eqref{SWM}. The resulting prices are plotted in Figure \ref{fig:v_synth}.
We can clearly see that the equilibrium price appears to increase linearly with increasing $k$, albeit the values of $a$ decrease and even become negative. This suggests that under this simplified setup, there may exist equilibria with arbitrarily high prices. However, considering network constraints and capacity constraints appears not to change this result, although $(m^+,a^+)$ are not guaranteed to be equilibrium strategies under \eqref{SWM}.  Nevertheless, this result requires complete information and assumes that players are willing to bid even very negative values of their $a_i$ parameter. This is rather unlikely in practice and although this reveals critical limitations of the SWM mechanism, the price is unlikely to become arbitrarily large as players may view bidding very negative values of $a_i$ as very risky, complete information is not available, and the multitude of equilibria provides another barrier to the realization of these outcomes. As an attempt to account for these issues, we introduced the Robust Strategy Selection approach.
\subsection{Real world case study of the CWE spot market}\label{sec:numRes}
The CWE market over the period we are concerned with (2019-2020), was composed of five zones (roughly) corresponding to five countries: Austria, Belgium, Germany (plus Luxembourg), France and the Netherlands. We evaluate the suitability of our models by comparing directly with the historical time series of prices for the corresponding time period, which we hereafter refer to as \textit{target prices}. The majority of computations are performed via \textit{Descartes Labs'} computing services \cite{desclab}.
\subsubsection{Input data}
We apply the Game-Theoretic model introduced here, and also solve \eqref{SWM} under the assumption of truthful bidding by solving independently for $2712$ consecutive hourly slots, with the first hour starting on the first hour of 
2 Jan 2019. Thus we require the temporal time series for player's characteristics $(c(t),b(t),Q(t))$, network constraints $\{M_p(t),b_p(t)\}$, and zonal demand levels $d^Z(t)$. The demand time series is obtained by using the day ahead forecasted values available at \cite{ENTSOEData}.
\par
Obtaining the time series of the full network constraints $\{M_p(t),b_p(t)\}_{t \in \mathcal{T}}$ set is a challenging task, and this is obtained by first obtaining the historical time series of the PTDFs and RAMs as described in our paper in \cite{PuiuHauser2021}. If we denote the set of line-scenario combinations \cite{PuiuHauser2021} that are tracked by the operator as $\mathcal{LS}$, the concatenation of all these constraints at any time $t$ gives us a form of network constraints as following
\begin{equation}
    r^{\mathcal{LS}} \leq PTDF_t^{\mathcal{LS}}(y(t) - d^Z(t)) \leq R_t^{\mathcal{LS}},
\end{equation}
and the constraint set can be re-written in the form used in \eqref{SWM} as
\begin{equation}\label{polytopeformation}
    M_p(t) = \begin{bmatrix} - PTDF^{\mathcal{LS}}_t \\ +PTDF^{\mathcal{LS}}_t \end{bmatrix}; \hspace{1mm} b_p(t) = \begin{bmatrix} - r^{\mathcal{LS}}_t  - PTDF^{\mathcal{LS}}_td(t)\\ +R^{\mathcal{LS}}_t + PTDF^{\mathcal{LS}}_td(t) \end{bmatrix}.
\end{equation}
To ensure that the feasible domain is a (closed) polytope, we further add to this set box constraints defined by  $(1-\Delta_{max})d_z(t) \leq y_z(t) \leq (1-\Delta_{max}) d_z(t)$, $\forall z$, with $\Delta_{max}=0.6$.
\par 
Before we obtain producer's characteristics at each time $t$, we must define the player set. For simplicity, similar to the work in \cite{mihaRH2015}, we define each player to represent all power plants of a specific type of production in one zone (e.g. one player for all Gas plants in Austria). We consider all main production types as per the \textit{Installed Capacity per Production Type} data available at ENTSOE \cite{ENTSOEData}. This gives around 11 players per zone, and 55 players in total. However, some of these players have very small capacity fraction compared to the total capacity in that country, and therefore they are unlikely to significantly affect the market outcome, but significantly increase computational cost. Since we require solving our game-theoretic model for thousands of time indexes and the network constraints have over $600$ rows, the computational cost can grow very large. As a result, we take a very parsimonious approach, and restrict the player set as an attempt to minimise computational cost whilst still retaining the dominant fundamental factors
\begin{figure}[ht!]
    \centering
    \includegraphics[width = 0.94 \columnwidth]{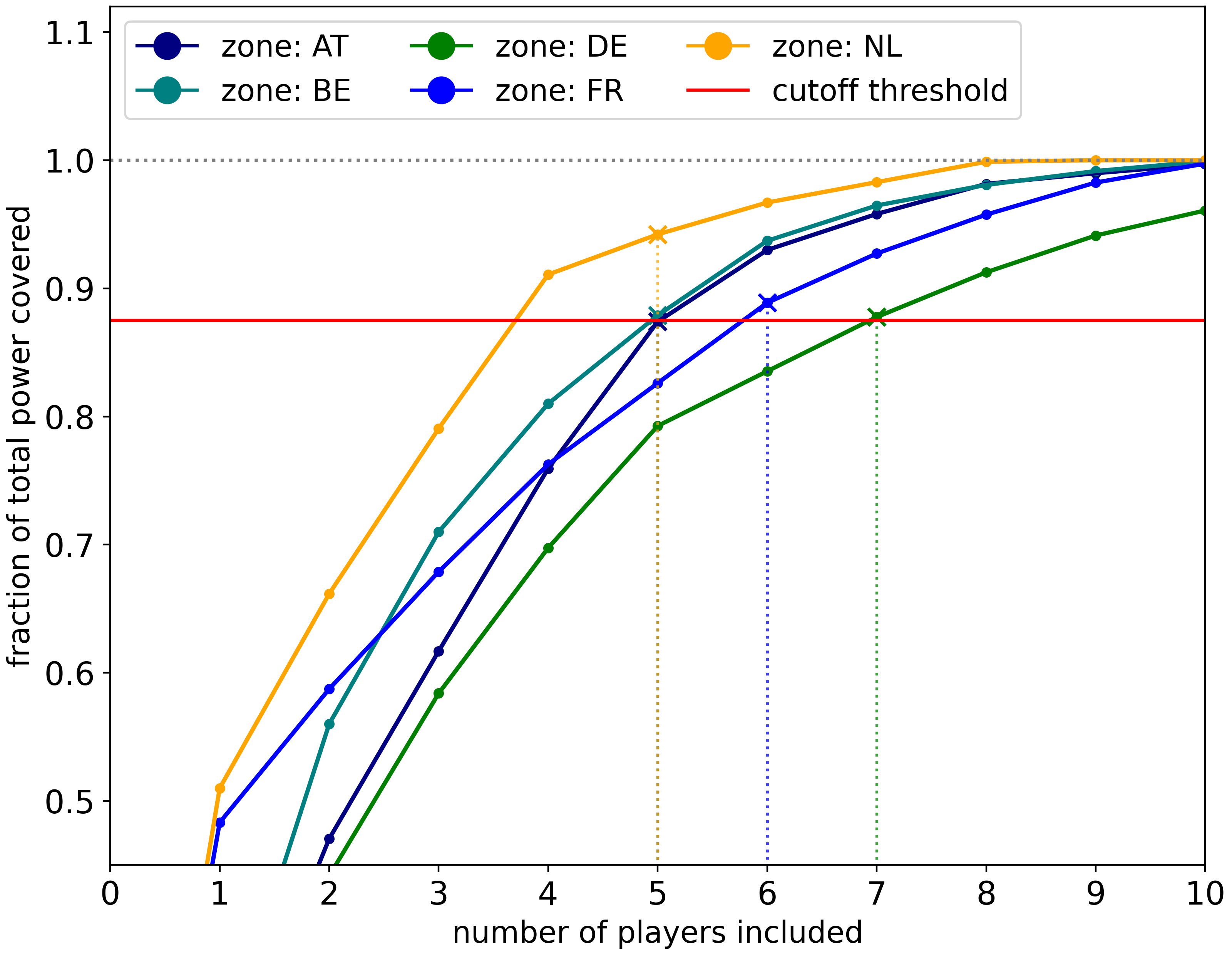}
    \caption{Thresholding process for retaining the top $n_z^*$ players in each zone (country).}
    \label{fig:threshold}
\end{figure}
\begin{figure*}[ht!]
    \centering
    \includegraphics[width = 1.925 \columnwidth]{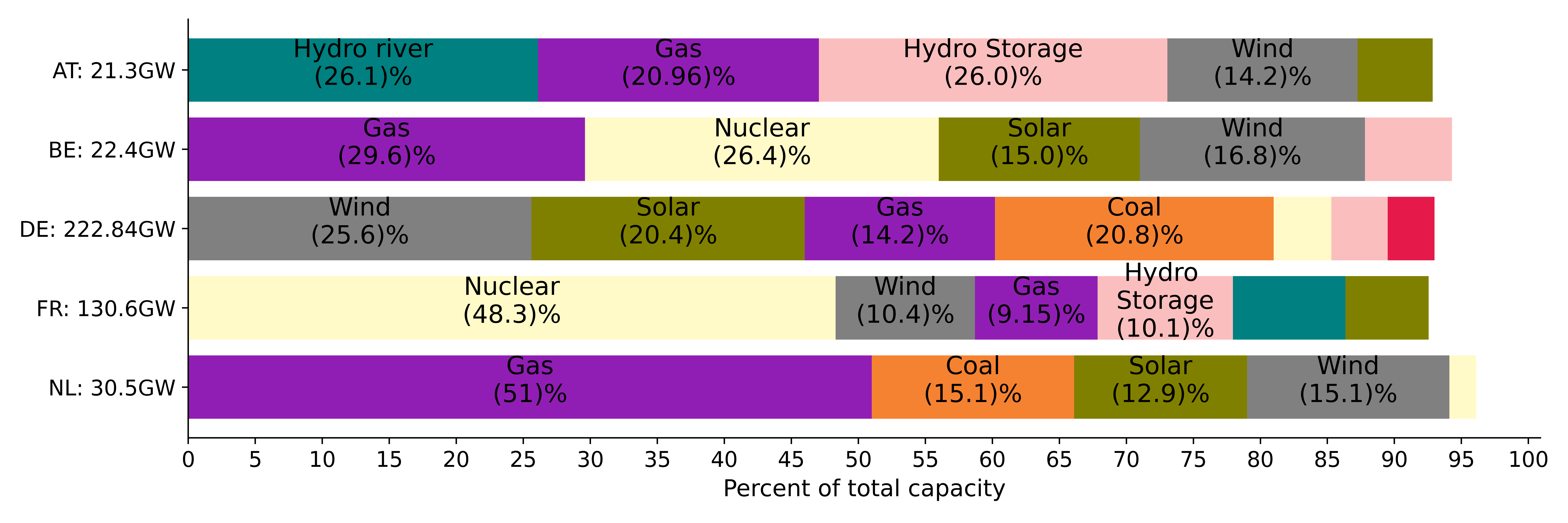}
    \caption{Player types and fraction of capacities covered for each country. The total capacity for each country is shown at the left hand side of the graph.}
    \label{fig_perc}
\end{figure*}
For each zone we apply a thresholding approach and choose the top $n_z$ players covering a fraction of at least $88\%$ of the total capacity, that is, 
$n_z^*= \min\{n_z \hspace{0.1mm}: \hspace{1mm} \phi_z(n_z)\geq 0.88, n_z \geq 5, \hspace{1mm}, \hspace{1mm} n_z \in \mathbb{N}\}$, where $\phi_z(n_z)$ is the fraction covered for the top $n_z$ players, and we also choose at least five players to retain most of the diversity of production types. This thresholding approach is summarised in Figure \ref{fig:threshold}. As a final step, once $n_z^*$ is obtained, we combine similar players in one type where possible to increase the capacity covered but retain low player number. To this end we combine On- with Off-shore wind, Hydro Pumped Storage with Hydro Water Reservoir, and  Hard Coal with Brown Coal (Lignite). In this way, we get $28$, players, just over half of what we had initially, greatly reducing the computational cost. Their types and capacity fractions are shown Figure \ref{fig_perc}.
\par 
With the player set well defined, we now have to obtain for each player $i$, the corresponding time series of cost structure $(c_i(t),b_i(t))$ and available capacity $Q_i(t)$, which may differ from the total installed capacity (especially for renewables). As per the relevant types in Figure \ref{fig_perc}, we construct these time series for the following types: (i) Fossil Gas, (ii) Coal, (iii) Nuclear, (iv) Solar, (v) Wind, (vi) Hydro Pumped Storage, (vii) Hydro Water Reservoir and (viii) Hydro Run-of-river and poundage. For simplicity, for each player, we obtain the two cost parameters for each time index $t$ by extrapolating from a representative value indicating the production cost per MWh. This usually (but not always) corresponds to the fuel cost for one MWh of production. Let this cost for type $\theta$ in zone $z$ at time $t$ be $k_{\theta,z}(t)$. Note that the double $(\theta,z)$ defines a player, say $j$. To obtain two parameters we then define the cost function for player $j$ as
\begin{equation}
    \lambda_{j,t}(x) := k_{\theta,z}(t)\left(1 - \frac{f_{\theta}}{1-f_{\theta}} n_{\theta}\right)+ n_{\theta}\frac{k_{\theta,z} }{(1-f_{\theta})Q_j}x = b_j+c_jx \,
\end{equation}
where $\lambda_{j,t}(x)$ is the price asked at quantity $x$, $f_{\theta}$ is the fraction of total capacity at which the price asked is exactly $k_{\theta,z}(t)$, and $n_{\theta}+1$ is the number of times the production cost is higher at full capacity when compared to the base average level, $ k_{\theta,z}(t)$. We take the cost to be $k_{\theta,z}(t)$ at $50\%$ capacity level, i.e. $f_{\theta} =0.5$ for all types for simplicity and $n_{\theta} \in [0,1]$, to guarantee that $b_j \geq 0$, with the exact value of $n_{\theta}$ depending on the fuel type. We are now left with obtaining $k_{\theta,z}(t)$ and $n_{\theta}$.
 \begin{table}[ht]
\begin{center}
\begin{tabular}{|c|c|c|c|c| }
 \hline
 {\small Type} & {\small Data}& $k_{\theta,z}$ &$n_.$ &{\small Conversion}\\
  & &  & &{\small comments}\\
 \hline
 {\small Gas} & {\small spot prices} & {\small time series} & $0.2$ & {\small DTH} \\
  & {\small TTF} \cite{PowerNextData} & {\small as per data} &  &  \\
  \hline
   & {\small dark}  & {\small time series} & $0.4$ & $\frac{\$}{\text{\officialeuro}} = 0.89$ \\ 
  {\small Coal}& {\small spread},   & {\small as per data} &  & $\frac{MWh}{ton} = \frac{20}{11}$ \\ 
  &  {\small proxy}&  &  & {\small DTH}  \\ 
  &  {\small of \cite{argus}} &  &  &\\ 
  \hline
   & {\small estimate} & &  & {\small may sell} \\
 {\small Nuclear}& {\small value }& $13.8\frac{\text{\officialeuro}}{MWh}$ &$0.8$ & {\small at up to} \\
 & & & &  {\small 50\% loss}\\
 \hline
 {\small Wind and} & {\small day ahead} & &  & {\small strongly}\\
 {\small Solar }& {\small generation } &$0.5\frac{\text{\officialeuro}}{MWh}$ & $0.05$ & {\small variable}\\
  & {\small forecast}   & &  &  $Q_i(t)$\\
   &\cite{ENTSOEData}  & &  &\\
   \hline
  {\small Hydro} & {\small base load} & {\small time series}&  & \\
  {\small Storage} & {\small electricity} &  {\small as per data}& $0.2$&\\
  & &{\small prices }& &\\
  \hline
  {\small Hydro run} & {\small estimate} &$8.45\frac{\text{\officialeuro}}{MWh}$ & $0.1$& {\small constant}\\
  {\small of river} &{\small  value} & & & {\small value}\\
  \hline
\end{tabular}
\end{center}
\caption{\label{Tab:costs_tab} Summary of data, parameters and assumptions used to estimate costs.}
\end{table}
How these are obtained is summarised in Table \ref{Tab:costs_tab}, but a more detailed description is available in out related paper \cite{PuiuHauser2022P1}, which uses the same data. Where $k_{\theta,z}$ is stated explicitly it means that it is a constant, while for the production types it is given by the time series described under the data column. Finally, Gas and Coal spot daily data is converted to hourly by assuming constant value during the day and this is denoted by DTH in the table. 
\par 
Admittedly, the approach used here to obtain time series of cost parameters for each player is rather simplistic. However, we use this approach to facilitate focusing on the formulation, analysis and solution of the market clearing and our game-theoretic model. Further, in subsection \ref{sec:calibration} we calibrate the cost parameters to improve our rather coarse guess. We show that with minimial fitting of only two parameters per zone (i.e. 10 in total) we obtain results that are very representative of the market, capturing well the main features of the actual price series, with test set errors of only $24.5\%$ and $25.2\%$ for the truthful bidding (TB) and game-theoretic (GT) models. The test set is a rather large set of $672$ hours (or 28 days), and these results demonstrate the robustness of our fundamental approach but also suggest that the reconstructed network constraints form our work in \cite{PuiuHauser2021} offer a satisfactory representation of the true constraints, at least for the application at hand.
\subsubsection{Model calibration}\label{sec:calibration}
Due to the complexity of the game-theoretic model, which has one extra layer of complexity when compared to the \eqref{SWM} solved under the truthful bidding model, calibrating the game-theoretic model is very difficult. We by-pass this problem by first calibrating the \eqref{SWM} model under the truthful bidding assumption which we call SWM-TB, and use these parameters for SWM-TB when this model is applied to the CWE case study. We then perform a simple one step adjustment of the SWM-TB parameters for the \eqref{SWM}-based game theoretic model (denoted as SWM-GT), and this is explained at the end of this subsection.
\par 
The calibration approach for the SWM-TB is very simple and is constrained to adjusting the scale of the costs for each zone $z$, by two scale parameters $(s_z^c,s_z^b)$  such that the target prices are represented well by the costs on some training data. This requires expressing the adjusted costs as
\begin{equation}
    c_i(t)^{final} = s_z^cc_i(t)^{initial}; \hspace{1mm} b_i(t)^{final} = s_z^bb_i(t)^{initial} \hspace{1mm}\forall i \in \mathcal{P}_z.
\end{equation}
By taking $s_z^c=1$ and $s_z^b=1$ $\forall z$ we would simply use the values obtained in the previous subsection. However, we optimise over these values to improve our initial guess of the costs, while allowing for unchanged inter-temporal variability. For this approach, the number of fitted parameters and computational costs are very low. We aim for a small number of parameters to fit because in principle, our fundamental model should be representative of the market without requiring any fitting, if the cost structure and other inputs are descriptive and estimated accurately. A further advantage of this approach is computational: $\partial v_z/\partial s^._z$ can be easily computed without knowing the set of players that set the price, i.e. $\{i:m_ix_i+a_i=v_{z_i} \}$. The optimisation problem has the objective 
\begin{equation}
    F(s) = \frac{1}{N_T}\sum_{t=1}^{N_T}\sum_{z}(v_{z,t} - P_{z,t})^2,
\end{equation}
where $v_{z,t}$  and $P_{z,t}$ are the model and target prices respectively, at zone $z$ for time $t$. Since the scale parameters $s = (s^c,s^b)$ are not dependent on time, we have a very strongly over-determined Nonlinear Least Squares Problem. If we now let
\begin{equation}
    F_t(s) = \sum_{z}(v_{z,t} - P_{z,t})^2,
\end{equation}
we then have that
\begin{equation}\label{gradrefeq1233}
    \frac{\partial F_t}{\partial s_z^c} = 2(v_{z,t} - P_{z,t})c_{k_z}x_{k_z}; \hspace{2mm} \frac{\partial F_t}{\partial s_z^b} = 2(v_{z,t} - P_{z,t})b_{k_z},
\end{equation}
which by summation, and collecting for all zones $z$, gives $\nabla_{s}F(s)$. The second derivatives can be similarly computed as 
\begin{equation}
    \frac{\partial^2 F_t}{\partial^2 s_z^c} = c_{k_z}^2x_{k_z}^2; \hspace{4mm} \frac{\partial^2 F_t}{\partial^2 s_z^b} = b_{k_z}^2; \hspace{4mm} \frac{\partial^2 F_t}{\partial s_z^b\partial s_z^c} = b_{k_z}c_{k_z}x_{k_z},
\end{equation}
with all other entries in the $\nabla^2_{s} F$ being 0. We choose to perform Gradient Descent over Newton's Method, with line-search as globalisation method since the former appears to obtain lower values of the objective function with better generalisations, although our investigation on this particular matter is limited. Let us call the obtained parameters $(s^c(TB),s^b(TB))$.
\par
The scale parameters for SWM-GT are fit by first using $(s^c(TB),s^b(TB))$ as input parameters to compute the output price series of our SWM-GT model for a subset of $\mathcal{T}_{GT}^0=\overline{24,672}$ time indices, and computing the average price ratio for each zone
\begin{equation}
    \hat{r}_z = \frac{1}{|\mathcal{T}_{GT}^0|}\sum_{t \in \mathcal{T}_{GT}^0} \frac{v_{z,t}^{SWM-GT}}{P_{z,t}}.
\end{equation}
The scale parameters $s_z^c(GT)$ for the SWM-GT models are then updated as $s_z^c(GT) \gets s_z^c(TB)/\hat{r}_z$, $\forall z$ only once, but we do not re-scale $s_z^b$, i.e. $s_z^b(GT) \gets s_z^b(TB)$. While this approach is very simplistic, we are not aiming to obtain the parameters that best represent the cost structure given the prices, but rather simply to improve on our rather coarse initial guess of costs.
\par
The aware reader might observe that by definition $s_z^c(GT) \neq s_z^c(TB)$ in general, and that this cannot be true in practice since the costs do not depend on the model choice. While this is a valid point, we are here interested in evaluating the representation capacity of both SWM-TB and SWM-GT, as well as inferring if and when strategic bidding takes place. Simply fitting the SWM-TB model and using these parameters for both models biases the results in favour of the SWM-TB model. This is especially true since strategic bidding may be assimilated into the truthful bidding model by implying higher costs. This is indeed what we observe in practice as $(s^c(TB),s^b(TB))$ are always well above one. While we are prepared to accept that our coarse estimation of costs is an under-statement, we cannot simply say that all the error in price estimation is due to the initial cost structure. For this reason, we aim for a very robust and simple calibration process, where we fit a limited number of scale parameters with a rather simplistic approach, and we have to allow for differences in scale parameters for SWM-GT against SWM-TB. In other words, we can only compare the case where players are bidding truthfully and the cost structure is the one given by $(s^c(TB),s^b(TB))$, with strategic bidding and cost structure given by $(s^c(GT),s^b(TB))$. This is due to the uncertainty in cost parameters, and this problem can be by-passed fully by obtaining very accurate cost estimates via a bottom up approach, but unfortunately this is outside the scope of this paper. Note that the approach in this paper remains valid if one obtains such accurate cost estimates.
\par 
 We next apply our SWM-TB and SWM-GT models to the real world case of CWE day ahead electricity market.
\subsubsection{Model performance}
We apply the SWM based truthful bidding model (SWM-TB) and the SWM based game-theoretic model (SWM-GT) to CWE data representing the first $2688$ consecutive hours starting on the first hour of 2 Jan 2019. We split the data in a train set $\mathcal{T}_{train}:= \{t \in \mathbb{N}: 24\geq t < 2040\}$ and a test set $\mathcal{T}_{train}:= \{t \in \mathbb{N}: 2040 \geq t < 2712\}$, where $t$ is a hourly time index. We start with a high level comparison of the SWM-TB and SWM-GT models by summarising the relative mean error for both the train set $\mathcal{T}_{train}$ and test set  $\mathcal{T}_{test}$, in Table \ref{tab:GTvsTB}.
\begin{table}[ht]
\begin{center}
\begin{tabular}{|c|c|c|c|c|}
    \hline
     & {\small SWM-TB }& {\small SWM-TB}  & {\small SWM-GT} &{\small SWM-GT}  \\
     & ($\mathcal{T}_{train}$) & ($\mathcal{T}_{test}$) & ($\mathcal{T}_{train}$) & ($\mathcal{T}_{test}$) \\
     \hline
     {\small AT}& $20.3\%$ & $28.8\%$ & $21.9\%$ & $26.7\%$\\
     \hline
     {\small BE} & $14.2\%$ & $22.8\%$ & $18.5\%$ & $25.4\%$\\
     \hline
     {\small DE} & $26.1\%$ & $30.5\%$ & $30.0\%$ & $34.2\%$ \\
     \hline
     {\small FR} & $14.5\%$ & $19.6\%$ & $18.2\%$ & $17.6\%$ \\
     \hline
     {\small NL} & $13.6\%$ & $18.5\%$ & $17.4\%$ & $17.6\%$ \\
     \hline
     {\small AVG} & $17.9\%$ & $24.6\%$ & $21.4\%$  & $25.2\%$  \\
     \hline
\end{tabular}
\end{center}
\caption{\label{tab:GTvsTB} Comparison of relative mean errors for the Truthful Bids and Game Theoretic models, on train and test data sets. Row AVG denotes the aggregate results across all zones.}
\end{table}
 By direct comparison of the relative mean error, we observe that the Truthful Bids model outperforms the Game Theoretic one. However, the difference is rather small, and it is worth recalling that while the cost scale parameters for the SWM-TB were fit via Gradient Descent, only a one step heuristic fit was performed for the parameters of the SWM-GT model. Further, the summary statistic presented here does not reveal any details with regards to the temporal structure, and producers may choose to bid strategically only at specific times. A potential reason for this is that the regulators require the producers to bid truthfully, in an attempt to minimise market power exercised by the supply side, but producers are concerned with their own profit, and thus participants may attempt to conceal exercising their market power by bidding strategically only at specific times, with better reward-to-risk profile.

\begin{figure*}[ht!]
    \centering
    \includegraphics[width =15.75cm]{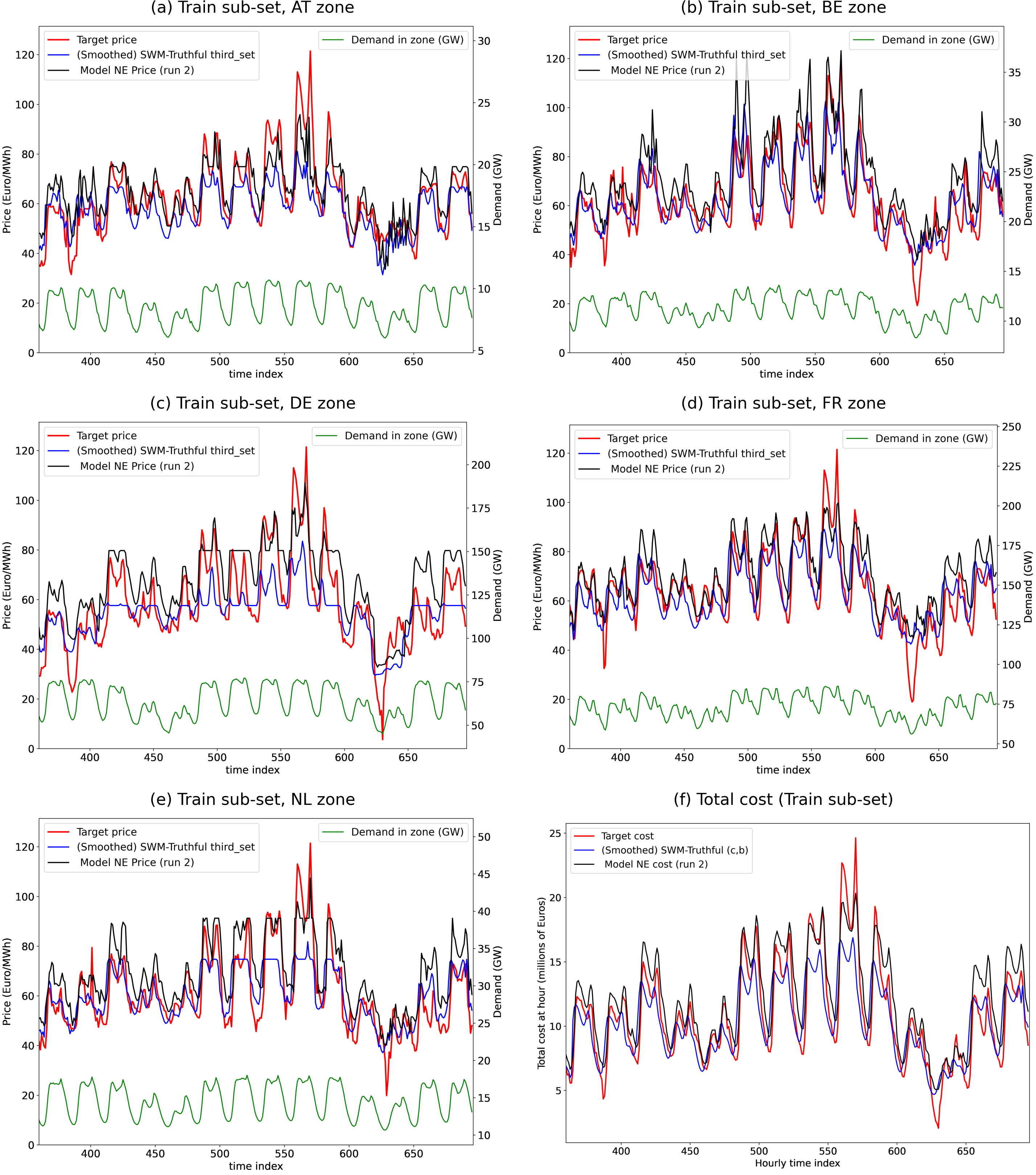}
    \caption{Price time series results (Train set): SWM - Truthful Bids (blue) and SWM - Game Theoretic (black) comparison against observed prices (red).}
    \label{fig:R3.1.train}
\end{figure*}

 A detailed view of the model and observed prices time series is shown in Figure \ref{fig:R3.1.train}, for a better intuition of models' performance. 
We can see that the Game Theoretic model still accurately traces the target price, but while the Truthful Bids model has the tendency to under-predict, the Game Theoretic one is more likely to over-predict. We can qualitatively observe that the GT model is often very accurate at peak hours, when compared to the TB one. This is rather remarkable, given the extremely simple calibration of the GT model, and suggests that players may not only bid strategically, but also strategically choose the times at which they do so. One may argue against the possibility of strategic bidding, on the premise that no parameter fitting is required for our models, since they are of fundamental nature. 
\begin{figure*}[ht!]
    \centering
    \includegraphics[width =15.75cm]{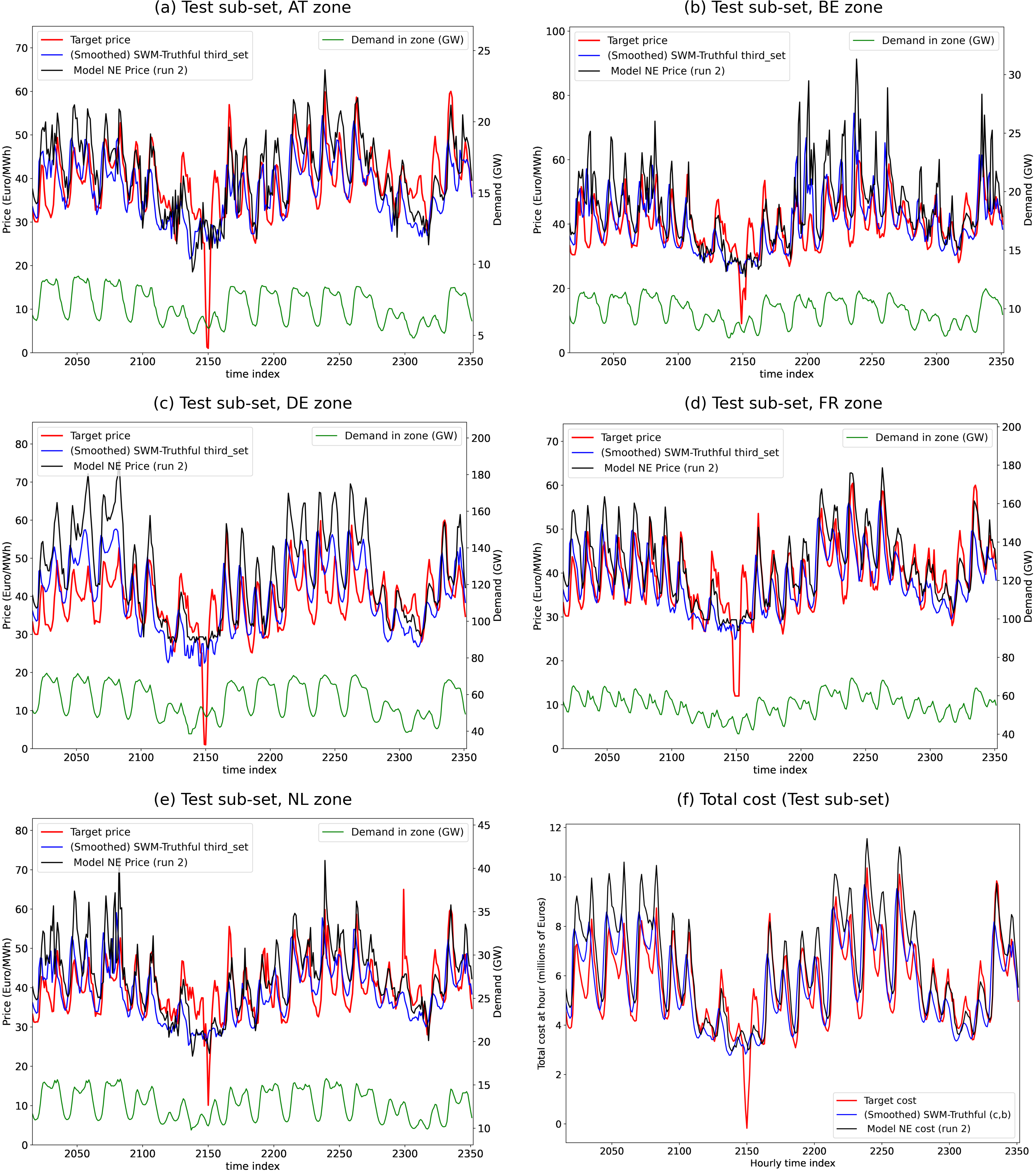}
    \caption{Price time series results (Test set): SWM - Truthful Bids (blue) and SWM - Game Theoretic (black) comparison against observed prices (red).}
    \label{fig:R3.1.test}
\end{figure*}
However, the complexity of the players' cost structure makes it very difficult to obtain accurate representations without very granular data and analysis, or calibration. Further note that calibration of the SWM-TB model to the observed price data could in fact adversely incorporate and conceal gaming. To observe this, imagine a tacit agreement between all players to bid exactly twice their costs. In this case, the value of our cost scales would be double, but the SWM-TB model would perfectly represent the price series. Indeed, for our fit, we observe our cost scales to be well above one. This issue can be best solved by having highly granular and reliable cost data and an accurate cost model. In our case, given the limited information on cost structure, it is precisely the inconsistencies in the bidding patterns that help us reveal strategic behaviour. We can observe that there are isolated days where prices are much higher, while the demand pattern is essentially the same. It is possible but unlikely for the production costs to change this fast, and thus assuming that the network constraints remain constant, the only other possibility is a state change: from truthful bidding to strategic bidding. 
\par 
Admittedly the price spikes could occur due to fast changes in network constraints, while our reconstructed data from \cite{PuiuHauser2021} may change more slowly. However, since we only use the reconstructed constraints data, without forecasting into the future, and given the results in Chapter 2, it seems reasonable to assume that our constraints time series captures the main features of the real one. From Figure \ref{fig:R3.1.test} we observe that both models perform very well on the test set, revealing the robustness of our fundamental models. 
\begin{figure*}[ht!]
    \centering
    \includegraphics[width =14.7cm]{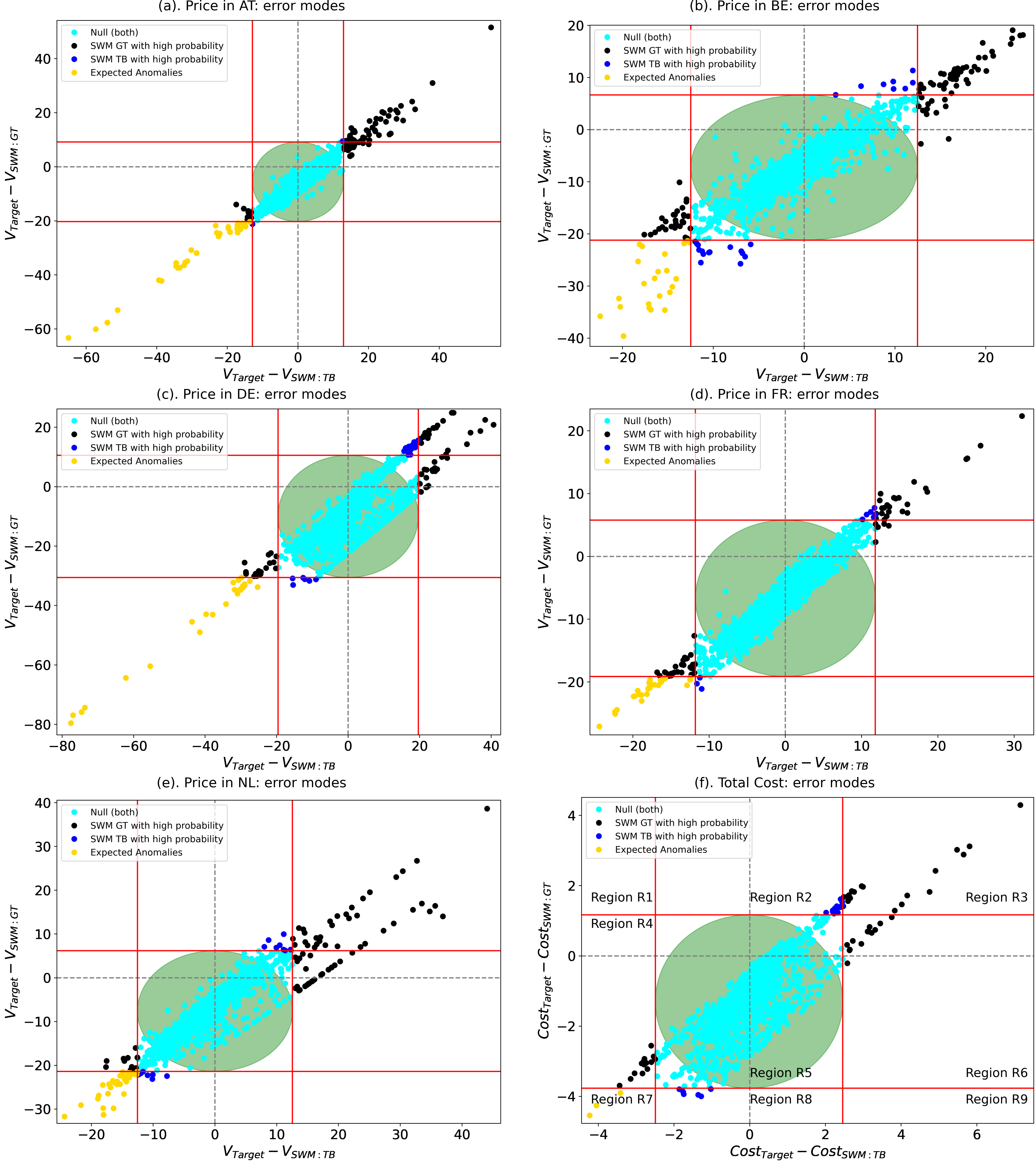}
    \caption{Classification of time points to their the most representative state. The red bars represent $97.5\%$ CI for each axis respectively.}
    \label{fig:R3.4.clusters}
\end{figure*}
\subsubsection{Uncovering strategic behaviour}
We next identify if and when players are gaming, assuming that our models, cost stucture and values, constraints and demand inputs are representative. We achieve this via simple hypothesis testing, and classifying different possible outcomes. Our analysis is performed based on price level, for each zone and per time point individually, and no time-dependent structure is considered. To this end we define two error time series, one for the TB and other for the GT models:
\begin{equation}
    e^{TB}_{z,t} = P_{z,t} - v_{z,t}^{SWM-TB}; \hspace{2mm} e^{GT}_{z,t} = P_{z,t} - v_{z,t}^{SWM-GT},
\end{equation}
where $P_{z,t}$ is the zonal price in zone $z$ at time $t$ and $v_{z,t}^{SWM-TB}$ and $v_{z,t}^{SWM-GT}$ are the zonal prices given by the SWM-TB and SWM-GT models respectively.
\par 
For each zone $z$, each time point is assigned the coordinates $\left(e^{TB}_{z,t},e^{GT}_{z,t}\right)$, and hypothesis testing is performed for each of the two error components individually. The null hypotheses are that the error at time $t$ comes from the error distribution of the TB (for $e^{TB}_{z,t}$) and GT (for $[e^{GT}_{z,t}$) models respectively, which are approximated by Gaussians with parameters set as their empirical estimators. Let us denote the null hypotheses corresponding to TB and GT as $\mathcal{N}_{TB}^0$ and$\mathcal{N}_{GT}^0$ respectively. We reject the null hypothesis if the corresponding error (for TB or GT) lies outside a symmetric $97.5\%$ confidence interval (CI) centred in the mean of the distribution. However, the null hypothesis can be rejected because $e^{.}_{z,t}$ is either \textit{"very low"} or \textit{"very high"}, and this gives us different information about the actual state of market. Thus, there are three regions of interest when performing hypothesis testing for each of the TB and GT models. Since we are interested in the cartesian product of the $e^{TB}$ and $e^{GT}$ space, this means that there are $9$ regions in total, and these are annotated in
Figure \ref{fig:R3.4.clusters} (f), for visualisation. If the error lies in the low end then the model over-predicts while a high error indicates under-predicting. The meaning of each region is then as following:
\begin{enumerate}
    \item If a time point lies in region R5 then neither $\mathcal{N}_{TB}^0$ nor $\mathcal{N}_{GT}^0$ can be rejected, and we cannot confidently say if players were bidding truthfully or strategically. We call this state \textit{Null} and denote it as $\mathcal{S}_0$.
    \item If a time point lies in R4 or R6, then $\mathcal{N}_{TB}^0$ is rejected, but not $\mathcal{N}_{GT}^0$, meaning that \textit{with high probability} (w.h.p), the players were bidding strategically. We call this state \textit{GT (w.h.p)} and denote it as $\mathcal{S}_{GT}$.
    \item If a time point lies in R2 or R8, then $\mathcal{N}_{GT}^0$ is rejected, but not $\mathcal{N}_{TB}^0$, meaning that w.h.p the players were bidding truthfully. We call this state \textit{TB (w.h.p)} and denote it as $\mathcal{S}_{TB}$.
    \item If a time point lies in R3, both models under-predict with respect to the observed price. This suggests that both models are inappropriate in this case. However, under the assumption of accurate inputs in our model, the only possibility for this to happen is if the players bid strategically, although different risk aversions or game setups may be relevant in this case. We also assign these points to state \textit{GT (w.h.p)}.
    \item If a time point lies in R7, then both models over-predict, meaning that the actual price is much lower. This case seems to occur when electricity prices become negative due to unexpected and high influx of renewable power, and thus we call these points \textit{Expected Anomalies} and denote their set as $\mathcal{S}_{EA}$.
    \item It is extremely unlikely for a point to lie in R1, as this means that the GT model severely under-predicts, while the TB model severely over-predicts. This requires the TB model giving a much higher price than the GT model. Due to multiple zones, this is not necessarily impossible but extremely unlikely. It is also fairly unlikely for a point to lie in R9 where the GT model drastically over-predicts, while the TB model drastically under-predicts. We do not observe any of these points on our data set but classify these points as \textit{Other Anomalies} and denote their set as $\mathcal{S}_{OA}$.
\end{enumerate}
\par 
\begin{figure*}[ht!]
    \centering
    \includegraphics[width = 15.75cm]{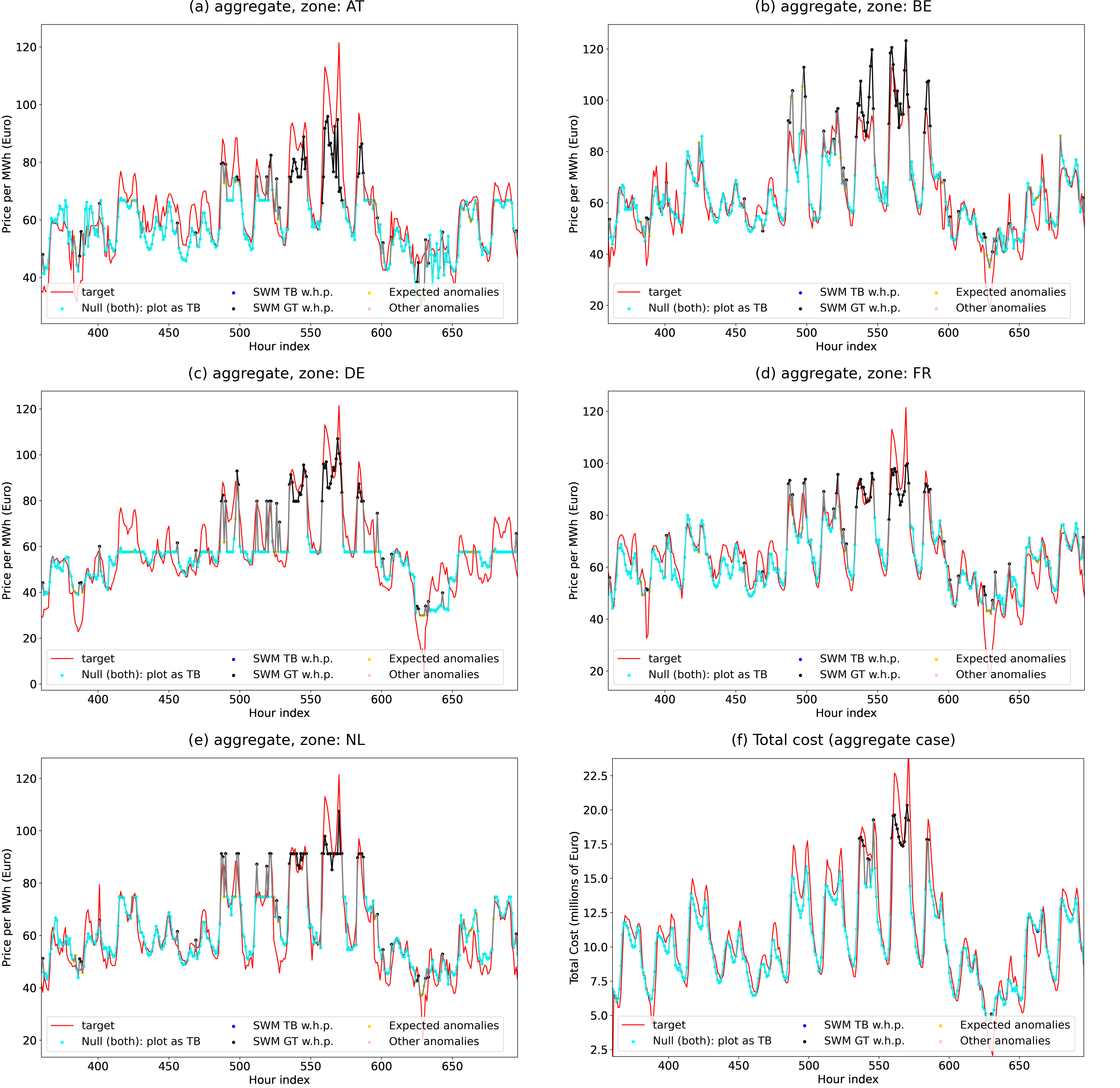}
    \caption{Comparison of price time series (aggregate TD-SD):  State Detection results vs observed prices}
    \label{fig:R3.5.i}
\end{figure*}
We call the method of assigning these states Two Dimensional State Detection (TD-SD).
While the choice of confidence threshold may impact results, this simple model enables us to gather some insight into whether the players were bidding truthfully or strategically. Since the analysis was performed on the price time series for each zone individually, we next aggregate the results for a final single state at each time $t$ over all $z$ based on the results obtained for each of the $5$ zones, and this requires further definitions. We found that performing the same hypothesis testing on total cost, the resulting price series is less accurate, because price decoupling happens fairly often and thus gaming may occur only in one or few regions, but this becomes harder to identify at market level. For this reason, we perform TD-SD independently for each zone, and define logical rules for aggregation. Let $\varsigma(t) \in \mathbb{R}^{N_z}$ be the vector of states assigned for each state, then we define the aggregate state $\chi_t$ as following:
\begin{enumerate}
    \item If $\varsigma_z(t) \in \mathcal{S}_0$ $\forall z$, then the aggregate state is $\chi_t \in \mathcal{S}_0$. 
    \item If $\varsigma_z(t) \in \mathcal{S}_0$ $\forall z$, $\exists z$ s.t. $\varsigma_z(t) \in \mathcal{S}_{TB}$ and $\nexists z$ s.t. $\varsigma_z(t) \in \mathcal{S}_{GT} \cup \mathcal{S}_{EA} \cup \mathcal{S}_{OA}$, then $\chi_t \in \mathcal{S}_{TB}$. 
    \item If $\exists z$ s.t. $\varsigma_z(t) \in \mathcal{S}_{GT}$ and $\nexists z$ s.t. $\varsigma_z(t) \in \mathcal{S}_{EA} \cup \mathcal{S}_{OA}$, then $\chi_t \in \mathcal{S}_{GT}$.
    \item If $\exists z$ s.t. $\varsigma_z(t) \in \mathcal{S}_{EA}$ and $\nexists z$ s.t. $\varsigma_z(t) \in \mathcal{S}_{OA}$, then $\chi_t \in \mathcal{S}_{EA}$.
    \item If $\exists z$ s.t. $\varsigma_z(t) \in \mathcal{S}_{OA}$, then $\chi_t \in \mathcal{S}_{OA}$.
\end{enumerate}
While the aggregation rules are mostly a natural extension, there is some degree of arbitrariness. However, the aggregation of zonal results enable us to summarise the main observations about the market as a whole, instead of describing each country individually (and their interaction), which is much more complex. Nevertheless, the zonal results can also be used if a more detailed analysis is required. We present the price estimation based on the aggregated modes for the market in Figure \ref{fig:R3.5.i}.
\par 
Our approach gives an indication of when players are bidding strategically, and we can observe that times tend to be consecutive, usually at peak hours, and under consecutive days. Strategic bidding around peak hours could be an approach employed by margin producers to mitigate the risk of being detected at least with respect to potential gain, since producers' market power is at its peak when demand is highest.
\begin{figure*}[ht!]
    \centering
    \includegraphics[width=15.5cm]{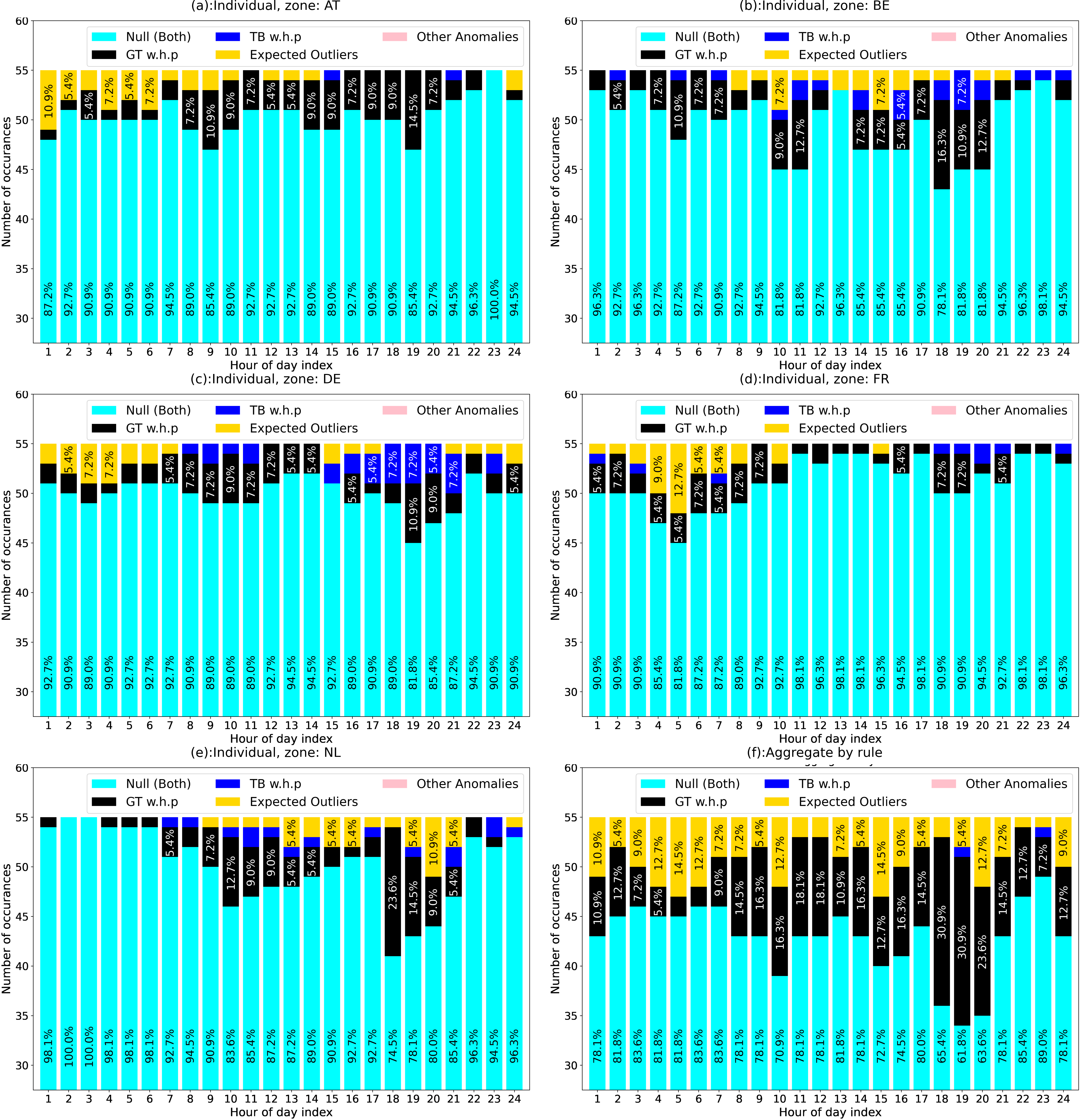}
    \caption{Results for the average day: Proportion of each observed state}
    \label{fig:R3.6}
\end{figure*}
\par Since it is very difficult to visualise the results for the whole length of the series, and ingest the main conclusions in this way, we review the state profile of an \textit{average day}. To this end, we count the number of each type of states for each hour of the day, and plot the results in Figure \ref{fig:R3.6}. We observe that generally, for each zone we can say with high confidence much more often that the players are bidding strategically, than we can say they are bidding truthfully. However, for a very large proportion of times we essentially cannot draw any conclusion.  This is due to the fairly large variance of our model errors, which is dependent on model structure and inputs. This could be potentially improved if data of higher quality and granularity is available, as well as by including more market features in the model. However, this requires more reliable data sources, extensive added labour and increased computational cost, and for these reasons we confined ourselves to the current approach. Reducing the confidence requirement will increase the fraction of times we can classify, but this increases the chance of over-stating results. Thus, we adopt a conservative approach, requiring $97.5\%$ confidence, and this still reveals interesting features. Nevertheless, the final general conclusion appears to remain the same even when varying the confidence threshold: strategic bidding occurs a significant number of times with high probability.
\par 
Figure \ref{fig:R3.6} reveals that strategic bidding can be observed in each zone for a significant fraction of times, and this generally tends to be more pronounced at intervals around peak hours: $8am-1pm$ and $5pm-9pm$. As discussed before, this could be due to a much more attractive risk-reward profile of strategic bidding alternative for the producers. We also observe that strategic bidding seems most pronounced in the Netherlands, and Belgium and least pronounced in France and Germany. This could be because the Netherlands and Belgium are fairly small trading zones with a fairly small number of players and therefore market power can be exercised much more easily. Heavy reliance on gas power plants and electricity imports also aggravate this issue. On the other hand France is a large trading zone and much cheap production in the form of nuclear power plants is available. Further these plants cannot easily withhold capacity and thus they do not have strong incentives to bid strategically. Out of the five, Germany is the largest trading zone, with the largest number of participants, and a large number of coal power plants ready to produce if price increase due to strategic bidding exceeds the cost of extra emissions. As a result, we can observe that for Germany we obtain the largest number of times for which we can say with high confidence that the producers were bidding truthfully.
\par
Our results suggest that strategic bidding in Austria is also fairly elevated. While Austria is also a fairly small zone, the results could be somewhat over-stated due to valuation of Hydro storage power. This is difficult to price and our approach may be less conservative than the one employed by these plants' operations manager. 
\par 
We also observe a fairly pronounced number of times where both SWM-TB and SWM-GT models over-predict (Expected Outliers). This occurs mostly in early morning hours, when demand is low and renewable energy production may be unexpectedly high due to high wind. This appears to be most pronounced in France, which is expected since nuclear power plants are not flexible to adjust, and prefer to sell at lower prices or even a loss.
\par The aggregate results presented in Figure \ref{fig:R3.6} indicate that while we cannot identify with certainty the bidding behaviour of players for a large number of times, strategic bidding (in at least one zone at each time) may be fairly pronounced. In fact, the fraction of times we can conclude with high confidence that players in all zones were bidding truthfully is very small. While it is possible that all the unidentified time instances truly belong to the truthful bidding case, even in this situation, our results reveal that the fraction of strategic bidding may be much higher than what market participants, regulators and academic papers generally assume. 
\section{Conclusion}
We considered the case of strategic bidding in spot electricity markets, focusing on the market mechanism used in European markets, which to our knowledge was not studied before. We take the demand as inflexible, network and capacity constraints are considered, while submitted orders are restricted to supply functions with linearly increasing price with respect to the produced quantity. We present theoretical results on a simplified case revealing that infinitely many equilibria may exist, if players are risk neutral, and show numerically that equilibria with arbitrarily high prices may exist. As a solution, a novel type of non-parametric risk aversion based on a defined worst case scenario is introduced, and this ensures boundedness of equilibrium prices and reduces dimensionality of the strategy space. By leveraging these properties we devise Jacobi and Gauss-Seidel type of iterative schemes for computation of approximate global Nash Equilibria, which is in contrast to the derivative based local equilibria usually obtained by solving Equilibrium Problems with Equilibrum Constraints (EPECs).
\par 
For the case of strategic bidding (represented by the game theoretic framework), as well as truthful bidding we apply our resulting models to the real world data of Central Western European Day Ahead market during the 2019-2020 period. We show that both our models offer a good representation of the historical time series of prices. We finally devise a simple method based on hypothesis testing to infer if and when producers are bidding strategically, instead of truthfully, and we find evidence that strategic bidding may be fairly pronounced in the CWE region.
\newpage
\newpage
\appendix
\newpage
\newpage
\section{Proofs of theoretical results}
\label{sec:appendix_proofs}
\begin{nono-result}\label{unconsSWMresult}
Consider a one zone market with inflexible demand $d>0$ and producer set $\mathcal{P}$. Assume producers have no capacity constraints and that there are no network constraints. Assume that the submitted orders have $m_i>0$, $\forall i$, and that the market clearing takes the form \eqref{SWM-s0}. Then the optimal quantities allocated to active players in set $\mathcal{P}^A$ are
\begin{equation}\label{xivaluncons}
\begin{aligned}
    x_i^* &= \frac{d+ \sum_{j \in \mathcal{P}^A \setminus \{i\}}m_j^{-1}(a_j-a_i)}{m_i \sum_{j\in \mathcal{P}^A}\frac{1}{m_j}}, \hspace{2mm} i \in \mathcal{P}^A\\
\end{aligned},
\end{equation}
and $x_i=0$ if $i \notin \mathcal{P}^A$. Further, the marginal prices for any player in the active set $\mathcal{P}^A$ gives the zonal price, that is $v= \lambda_i=m_ix_i^*+a_i$.
\begin{proof}
Without loss of generality we can assume that $\mathcal{P}=\mathcal{P}^A$, since all other players are irrelevant. Problem \eqref{SWM-s0} can be simplified by re-writing the balance constraint as $x_n = d - \sum_{i \in \mathcal{P} \setminus \{n\}}x_i$ and plugging this in the objective to yield 
\begin{equation}
\begin{aligned}
    \min_{x_i:i\neq n}&\sum_{i\neq n}\left(\frac{1}{2}m_ix_i^2+a_ix_i\right)+ \frac{1}{2}m_n\left( d - \sum_{i \in \mathcal{P} \setminus \{n\}}x_i\right)^2 \\ &+a_n \left( d - \sum_{i \in \mathcal{P} \setminus \{n\}}x_i\right)
\end{aligned}\label{simplified_SWM-s}
\end{equation}
which is an unconstrained problem with one degree of freedom corresponding to $x_n$ removed. We can set the gradient of this objective to zero to obtain the minimum. This gives us the following system of equations
\begin{equation}\label{eq2appendix}
    m_ix_i+a_i = m_n\left( d - \sum_{i \in \mathcal{P} \setminus \{n\}}x_i\right) + a_n \hspace{3mm} \forall i,
\end{equation}
which tells us that at optimality the marginal price of all players with $x_i>0$ is the same, and thus $v=\lambda_i=m_ix_i+a_i$, $\forall i:x_i>0$. System \eqref{eq2appendix} can be solved by writing it as
\begin{equation}
    \left(\hat{D}_m + m_n\hat{e}\hat{e}^T\right)\hat{x} = m_n d \hat{e} + a_n\hat{e} - \hat{a}
\end{equation}
where $\hat{x}$, $\hat{a}$, $\hat{m}$ represent the vectors of quantities for all players except $n$, $\hat{e}$ is the vector of all ones of length $|\mathcal{P}|-1$, and $\hat{D}_m= diag(\hat{m})$. This system is full rank since $m_i>0$ $\forall i$, and an analytic solution can be obtained by applying Woodbury's matrix identity as an inversion formula which in this case leverages the diagonal structure of $\hat{D}_m$ and the rank one of $\hat{e} \hat{e}^T$.
\par 
The optimal value of $x_i$ for $i \neq n$ is then
\begin{equation}
    x_i = \frac{d+ \sum_{j \neq i}\frac{a_j-a_i}{m_j}}{m_i \sum_{j}\frac{1}{m_j}}.
\end{equation}
However, the above equation does not contain $n$ indices and there was no specific property of player $n$ used in the derivation. Thus, by symmetry the equation above must hold for player $n$ as well.
\end{proof}
\end{nono-result}

\begin{nono-theorem}\label{nono-unique_NE_in_A}
Consider a one zone market with inflexible demand $d>0$ and active producer set $\mathcal{P}$. Assume producers have no capacity constraints and that there are no network constraints. Further assume that demand $d$, cost structure $\{c_i,b_i\}_{i \in \mathcal{P}}$ and network constraints $(M_p,b_p)$ are common knowledge, and that each producer is aiming to maximise its profit, $\pi_i(m_i,a_i,m_{-i},a_{-i})=vx_i-\frac{1}{2}c_ix_i^2-b_ix_i$, with $v$ the zonal clearing price. Let the market clearing be given by \eqref{SWM-s0}. Finally, assume that for every $i$ we have $c_i>0$, $m_i^+\geq\frac{1}{2}c_i$ is held fixed and that $a_i$ is the only strategy variable, and that this is common knowledge. Then for any vector $m^+ \in \mathcal{M}$ there exists a unique Nash equilibrium $a^+(m^+) \in \mathcal{A}$, which is defined by the (full rank) linear system
\begin{equation}\label{linsysa}
    \theta_i\sum_{j \in \mathcal{P}\setminus\{i\}}m_j^{-1}a_i^+ - \sum_{j \in \mathcal{P}\setminus\{i\}}\frac{a_j^+}{m_j} = d + (\theta_i-1)\sum_{j \in \mathcal{P}\setminus\{i\}}\frac{b_j}{m_j},
\end{equation}
where $ \theta_i = \frac{2 - k_iK_{-i}^2}{1 - k_iK_{-i}^2}$, $k_i=2m_i-c_i$, $K_{-i}^2 = \frac{\sum_{j \in \mathcal{P}\setminus\{i\}}m_j^{-1}}{m_i\sum_{j \in \mathcal{P}}m_j^{-1}}$.
\begin{proof}
The proof is obtained by using the optimal form of $x_i$ values for the \eqref{SWM-s0} problem, as per Result \ref{unconsSWMresult}, given $\{m_i,a_i\}_{i \in \mathcal{P}}$. Using the expression for each $x_i$, the local Nash Equilibria equations are obtained, and the final part of the proof shows this is a full rank, square linear system. 
\par 
We now look at the profit maximisation problem of each player $i$. By using Result \ref{unconsSWMresult}, the profit function can be written as 
\begin{equation}
    \pi_i = vx_i-\frac{1}{2}c_ix_i^2-b_i= m_ix_i^2+a_ix_i -\frac{1}{2}c_ix_i^2-b_ix_i,
\end{equation}
and by taking the partial derivative with respect to $a_i$ we obtain
\begin{equation}
    \frac{\partial \pi_i}{\partial a_i} = (k_ix_i+a_i-b_i) \frac{\partial x_i}{\partial a_i} + x_i,
\end{equation}
where $k_i = 2m_i-c_i$. Using \eqref{xivaluncons}, we get that 
\begin{equation}
     \frac{\partial x_i}{\partial a_i} = -\frac{\sum_{j \in \mathcal{P}\setminus\{i\}}m_j^{-1}}{m_i\sum_{j \in \mathcal{P}}m_j^{-1}} =: -K_{-i}^2,
\end{equation}
and we can re-write $x_i$ as 
\begin{equation}
    x_i = K_{-i}^1 -  K_{-i}^2a_i,
\end{equation}
where
\begin{equation}
    K_{-i}^1 =\frac{d}{m_i\sum_{j \in \mathcal{P}}m_j^{-1}}.
\end{equation}
By using these results and imposing the first order condition that $\frac{\partial \pi_i}{\partial a_i} = 0$ we have that
\begin{equation}
    \frac{\partial \pi_i}{\partial a_i} = (1-k_iK_{-i}^2)x_i-K_{-i}^2(a_i-b_i)=0,
\end{equation}
which is equivalent to
\begin{equation}\label{linsysa}
    \theta_i\sum_{j \in \mathcal{P}\setminus\{i\}}m_j^{-1}a_i - \sum_{j \in \mathcal{P}\setminus\{i\}}\frac{a_j}{m_j} = d + (\theta_i-1)\sum_{j \in \mathcal{P}\setminus\{i\}}\frac{b_j}{m_j},
\end{equation}
where 
\begin{equation}
    \theta_i = \frac{2 - k_iK_{-i}^2}{1 - k_iK_{-i}^2}.
\end{equation}
This system is linear in $a$ given fixed $m$ and can be re-written in vector form as
\begin{equation}
    \left(\hat{D}^m+ev_m^T\right)a = \hat{r},
\end{equation}
where $\hat{D}^m$ is a diagonal matrix with $\hat{D}^m_{i,i}=\theta_i\sum_{j \in \mathcal{P}\setminus\{i\}}m_j^{-1}$, $v_m$ is a vector with $(v_m)_i = m_i^{-1}$ and $\hat{r}_i=d + (\theta_i-1)\sum_{j \in \mathcal{P}\setminus\{i\}}\frac{b_i}{m_j}$. Since $1+e^T\hat{D}^mv_m=1+ \sum_{j}m_j^{-1}>1\neq0$ by Sherman-Morrison formula, the matrix $\hat{D}^m+ev_m^T$ is indeed invertible meaning that system \eqref{linsysa} has a unique solution. Further, the exact inverse can be easily computed, albeit this is unnecessary here. Thus, there is a unique stationary point, and we next show that this is indeed a maximum for each player, by showing that $\frac{\partial^2 \pi_i}{\partial a_i^2}<0$. Differentiating we get that
\begin{equation}
\begin{aligned}
    \frac{\partial^2 \pi_i}{\partial a_i^2} &= 2 \frac{\partial x_i}{\partial a_i} + k_i \left(\frac{\partial x_i}{\partial a_i}\right)^2\\&= 2K_{-i}^2 \left(-1 + \frac{(1-\frac{c_i}{2m_i})\sum_{j \in \mathcal{P}\setminus \{i\}}m_j^{-1}}{\sum_{j \in \mathcal{P}}m_j^{-1}} \right).
\end{aligned}
\end{equation}
Since $K_{-i}^2>0$, $0<1-\frac{c_i}{2m_i}<1$ and $0<\frac{\sum_{j \in \mathcal{P}\setminus \{i\}}m_j^{-1}}{\sum_{j \in \mathcal{P}}m_j^{-1}}<1$ it follows that $\frac{\partial^2 \pi_i}{\partial a_i^2}<0$ concluding our proof.
\end{proof}
\end{nono-theorem}
\begin{nono-theorem}\label{nono-Theorem_inf_NE}
Consider identical conditions to Theorem \ref{unique_NE_in_A}, with the exception that $m_i$ are not held fixed, but rather part of the strategy choice for each player $i$. Then for each $m^+ \in \mathcal{M}$, there exists an $a^+(m^+)$, computed according to Theorem \ref{unique_NE_in_A} by taking $m^+$ as fixed, such that $(m^+,a^+)$ is a local Nash equilibrium in the $(m,a) \in \mathcal{M} \times \mathcal{A}$ space. In other words, since $m^+$ is chosen from an infinite set, there are infinitely many local equilibria. Further, the local equilibria form a continuous subspace $\mathcal{MA}^+$.
\begin{proof}
The idea of the proof has two steps. In the first step we show that any point $(m,a)$ satisfying $\frac{\partial \pi_i}{\partial a_i}=0$ also satisfies $\frac{\partial \pi_i}{\partial m_i}=0$, meaning that a stationary point for $\mathcal{A}$ space is also stationary for $\mathcal{M} \times \mathcal{A}$ space. The second part of the proof shows that the hessian of $\pi_i$ for each $i$, $\nabla^2\pi_i \in \mathbb{R}^{2 \times 2}$ has negative trace and zero determinant at $(m^+,a^+)$, meaning that the stationary point is a local (but not isolated) maximum for each player $i$, and that these lie on a one dimensional subspace for each player. 

We start by looking at $\frac{\partial \pi_i}{\partial m_i}$ and we get by differentiation of $x_i^*$ that
\begin{equation}
    \frac{\partial x_i}{\partial m_i} = -\sum_{j\neq i}\frac{1}{m_j}\frac{d+\sum_{j \neq i}\frac{a_j-a_i}{m_j}}{(m_i \sum_{j}\frac{1}{m_j})^2} =-K_{-i}^2x_i =  x_i\frac{\partial x_i}{\partial a_i}.
\end{equation}
We now have that
\begin{equation}
\begin{aligned}
    \frac{\partial \pi_i}{\partial m_i} &=  (k_ix_i+a_i-b_i) \frac{\partial x_i}{\partial m_i} + x_i^2\\
    &=(k_ix_i+a_i-b_i) \frac{\partial x_i}{\partial a_i}x_i + x_i^2\\
    &=x_i\left((k_ix_i+a_i-b_i) \frac{\partial x_i}{\partial a_i} + x_i\right)\\ &= x_i  \frac{\partial \pi_i}{\partial a_i}.
\end{aligned}
\end{equation}
Since this means $\frac{\partial \pi_i}{\partial a_i}=0 \implies \frac{\partial \pi_i}{\partial m_i}=0$, the first part of the proof is now done. Given any $m^+$, and obtaining $a^+(m^+)$ as per Theorem \ref{unique_NE_in_A}, $(m^+,a^+)$ satisfies the first order stationarity conditions. We next show that this stationary point gives a local maximum for each player $i$. From proof of Theorem \ref{unique_NE_in_A} we have that $\frac{\partial^2 \pi_i}{\partial a_i^2}<0$. Let us now evaluate the sign of $\frac{\partial^2 \pi_i}{\partial m_i^2}$ at $(m^+,a^+)$. We have that
\begin{equation}
    \frac{\partial \pi_i}{\partial m_i} =  x_i  \frac{\partial \pi_i}{\partial a_i},
\end{equation}
and therefore 
\begin{equation}\label{dmda_eq}
    \frac{\partial^2 \pi_i}{\partial m_i \partial a_i}=\frac{\partial}{\partial a_i}\left(\frac{\partial \pi_i}{ \partial m_i}\right)= \frac{\partial \pi_i}{\partial a_i} \frac{\partial x_i}{\partial a_i} + x_i\frac{\partial ^2\pi_i}{\partial a_i^2}.
\end{equation}
Then, using that $\frac{\partial x_i}{\partial m_i} = x_i \frac{\partial x_i}{\partial a_i}$, we get that
\begin{equation}
    \frac{\partial }{\partial m_i}\left(\frac{\partial \pi_i}{\partial m_i} \right) = \frac{\partial }{\partial m_i}\left(x_i\frac{\partial \pi_i}{\partial a_i} \right) = x_i \frac{\partial x_i}{\partial a_i}\frac{\partial \pi_i}{\partial a_i} + x_i\frac{\partial^2 \pi_i}{\partial m_i\partial a_i},
\end{equation}
which can be re-arranged by using \eqref{dmda_eq} to obtain
\begin{equation}\label{dm2}
    \frac{\partial^2 \pi_i}{\partial m_i^2} =  2x_i \frac{\partial x_i}{\partial a_i}\frac{\partial \pi_i}{\partial a_i} + x_i^2\frac{\partial^2 \pi_i}{\partial a_i^2}.
\end{equation}
By definition of $(m^+,a^+)$, we have that $\frac{\partial \pi_i}{\partial a_i}=0$ at $(m^+,a^+)$ and therefore 
\begin{equation}
     \frac{\partial^2 \pi_i}{\partial m_i^2} =  x_i^2\frac{\partial^2 \pi_i}{\partial a_i^2} <0,
\end{equation}
at $(m^+,a^+)$. The trace of the hessian $\nabla^2 \pi_i$ is then 
\begin{equation}
    Tr(\nabla^2 \pi_i) =  \frac{\partial^2 \pi_i}{\partial m_i^2} +  \frac{\partial^2 \pi_i}{\partial a_i^2}<0
\end{equation}
at the equilibrium point  $(m^+,a^+)$. The determinant at this point is
\begin{equation}
    \begin{aligned}
    \left|\nabla^2 \pi_i\right| &=  \frac{\partial^2 \pi_i}{\partial m_i^2}  \frac{\partial^2 \pi_i}{\partial a_i^2} - \left(\frac{\partial^2 \pi_i}{\partial m_i \partial a_i} \right)^2\\
    &= \left(2x_i \frac{\partial x_i}{\partial a_i}\frac{\partial \pi_i}{\partial a_i} + x_i^2\frac{\partial^2 \pi_i}{\partial a_i^2} \right) \frac{\partial^2 \pi_i}{\partial a_i^2} - \left(\frac{\partial^2 \pi_i}{\partial m_i \partial a_i} \right)^2\\
    &= x_i^2\left(\frac{\partial^2 \pi_i}{\partial a_i^2}\right) - \left(0+x_i\frac{\partial^2 \pi_i}{\partial a_i^2}\right)^2\\
    &=0
    \end{aligned}
\end{equation}
where the second line follows from equation \eqref{dm2}, the third line follows by using equation \eqref{dmda_eq} and that $\frac{\partial \pi_i}{\partial a_i}=0$ at $(m^+,a^+)$. Simple linear algebra tells us that the two eigenvalues of $\nabla^2\pi_i$ satisfy
\begin{equation}
    \begin{aligned}
    \lambda_1 \lambda_2 &=   \left|\nabla^2 \pi_i\right| = 0\\
    \lambda_1 + \lambda_2 &=Tr(\nabla^2 \pi_i)<0
    \end{aligned}\hspace{1mm},
\end{equation}
which can only happen if $\lambda_1<0$ and $\lambda_2=0$ (w.l.o.g. we take $\lambda_1 \leq \lambda_2$. This tells us that the stationary point  $(m^+,a^+)$ indeed gives a local maximum for each player $i$, meaning that $(m^+,a^+)$ is a local Nash Equilibrium. Further, since $\lambda_2=0$, unilateral deviations by changing both $m_i$ and $a_i$ are possible while still maintaining exactly the same profit level. 
\end{proof}
\end{nono-theorem}

\begin{nono-result}\label{nono-result_non_neg}
Under the market clearing given by \eqref{SWM}, player $i$ is guaranteed non-negative profit by playing any $\sigma_i \in \mathcal{S}_R'(i):=\{(m_i,a_i):m_i \geq \frac{1}{2}c_i, \hspace{1mm} a_i \geq b_i \}$, for any $(m_{-i}, a_{-i})$.
\begin{proof}
If the player plays strategy $\sigma_i \in \mathcal{S}_R'(i) $ and he is out of the market ($x_i=0$), then $\pi_i=0 \geq 0$. Let us consider the case of $x_i>0$. By definition of $v_{z_i}=\max\{m_jx_j+a_j:x_j>0; j \in \mathcal{P}_{z_i} \}$ and since $i \in \mathcal{P}_{z_i}$, $x_i>0$ we have that
\begin{equation}\label{eqv_ineq}
    v_{z_i} \geq m_ix_i+a_i
\end{equation}
The profit of player $i$ can then be lower bounded as
\begin{equation}
    \pi_i = v_{z_i}x_i - \frac{1}{2}c_ix_i^2-b_ix_i \geq \left[(2m_i-c_i)x_i+(a_i-b_i) \right]x_i
\end{equation}
where the inequality comes by using \eqref{eqv_ineq}. Since $x_i \geq 0$ is guaranteed by \eqref{SWM}, and $m_i \geq \frac{1}{2}c_i$, $a_i \geq b_i$ by the fact that $(m_i,a_i) \in \mathcal{S}_R'(i)$, we conclude that $\pi_i \geq 0$. The choice of $m_{-i}$ and $a_{-i}$ is absolutely irrelevant for the argument, and thus this must be the case for any such values.
\end{proof}
\end{nono-result}

\section*{Acknowledgements}
We would like to thank CGM London Power and Gas desk within Macquarie Group for partially funding this research, and providing computational resources via \textit{Descartes Labs}. \textit{Ioan Alexandru Puiu} would like to further thank \textit{Vincent Guffens} for his support, \textit{Descartes Labs} for extending the account, and \textit{Jeremy Malczyk} and \textit{Chris Moulton} for offering technical support with the Descartes Labs platform.
 \bibliographystyle{elsarticle-num} 
 \bibliography{main}





\end{document}